# Gaussian-Chain Filters for Heavy-Tailed Noise with Application to Detecting Big Buyers and Big Sellers in Stock Market

Li-Xin Wang

*Abstract*—In this paper we propose a new heavy-tailed distribution --- Gaussian-Chain (GC) distribution, which is inspirited by the hierarchical structures prevailing in social organizations. We determine the mean, variance and kurtosis of the Gaussian-Chain distribution to show its heavy-tailed property, and compute the tail distribution table to give specific numbers showing how heavy is the heavy-tails. To filter out the heavy-tailed noise, we construct two filters --- 2$^{nd}$ and 3$^{rd}$-order GC filters --- based on the maximum likelihood principle. Simulation results show that the GC filters perform much better than the benchmark least-squares algorithm when the noise is heavy-tail distributed. Using the GC filters, we propose a trading strategy, named Ride-the-Mood, to follow the "mood" of the market by detecting the actions of the big buyers and the big sellers in the market based on the noisy, heavy-tailed price data. Application of the Ride-the-Mood strategy to five blue-chip Hong Kong stocks over the recent two-year period from April 2, 2012 to March 31, 2014 shows that their returns are higher than the returns of the benchmark Buy-and-Hold strategy and the Hang Seng Index Fund.

## I. Introduction

A basic stylized fact about asset returns is their heavy-tailed distribution ([18], [7]), i.e. large returns (positive or negative) are much more frequent than predicted by the Gaussian distribution. Many distribution functions have been proposed in the literature to model the heavy-tailed distributions, such as stable Lévy distribution [18], lognormal-normal distribution [8], Student distribution [5], hyperbolic distribution [11], normal inverse Gaussian distribution [4], discrete mixture of normal distributions [14], stretched exponential distribution [15], exponentially truncated stable distribution [10], and several others. Although these distribution functions could fit the particular empirical data sets very well, it is far less clear what are the mechanisms behind the data that generate these heavy-tailed distributions [24]. For example, it was suggested that asset returns are self-similar in different time scales ([18], [19]) which results in the stable Lévy distribution, but careful examination of more real data showed that asset returns are in general not scale invariant ([6], [10]). Since asset prices are determined by human actions (buying and selling), it is important to explore what kind of human actions or social structures generate the heavy-tailed return distributions. In this paper, we propose a new type of distribution --- Gaussian-Chain (GC) distribution, which tries to model the hierarchical structures that are prevalent in social organizations. We will prove that the Gaussian-Chain distributions are heavy-tailed, indicating that hierarchical structuring is one mechanism for generating the heavy-tailed distributions.

The Gaussian distribution is suitable for the cases where a large number of independent elements are added together, but if the agents are connected sequentially in a hierarchical fashion, what type of distribution will emerge? Consider the schematic example illustrated in Fig. 1. The CEO of a company needs an estimate of the price of a key material next year in order to make a production plan, and he decides to draw the estimate from the Gaussian distribution: $p_{t+1} \sim N(p_t, |\sigma^{(q)}|)$, where $p_t$ is the price of the material this year, $p_{t+1}$ is the price next year, and $N(p_t, |\sigma^{(q)}|)$ is Gaussian distribution with mean $p_t$ and standard deviation $|\sigma^{(q)}|$. But the problem of the CEO is that he does not know how to determine the parameter $\sigma^{(q)}$ that quantifies the uncertainty of his estimate, so he calls in his Head of Research and asks him to provide the value of $\sigma^{(q)}$. The Head of Research also uses the favorite Gaussian distribution to characterize the uncertainty parameter: $\sigma^{(q)} \sim N(0, |\sigma^{(q-1)}|)$, but his problem is again not knowing how to determine the parameter $\sigma^{(q-1)}$ that quantifies the uncertainty, so he passes the difficult to his Senior Researcher who, in the same spirit, uses the Gaussian model: $\sigma^{(q-1)} \sim N(0, |\sigma^{(q-2)}|)$, but still don't know how to determine $\sigma^{(q-2)}$. Finally, the Senior Researcher assigns the problem (determine $\sigma^{(q-2)}$) to his newly graduated Research Assistant who has no choice but to determine the parameter $\sigma^{(q-3)}$ in his model $\sigma^{(q-2)} \sim N(0, |\sigma^{(q-3)}|)$ by himself because he is at the bottom of the hierarchical chain and unable to pass the difficult to other person; he chooses $\sigma^{(q-3)} = \sigma$ where $\sigma$ is some non-random constant. Now the question is what the distribution of the CEO's estimates of next year's price $p_{t+1}$ looks like in this situation.

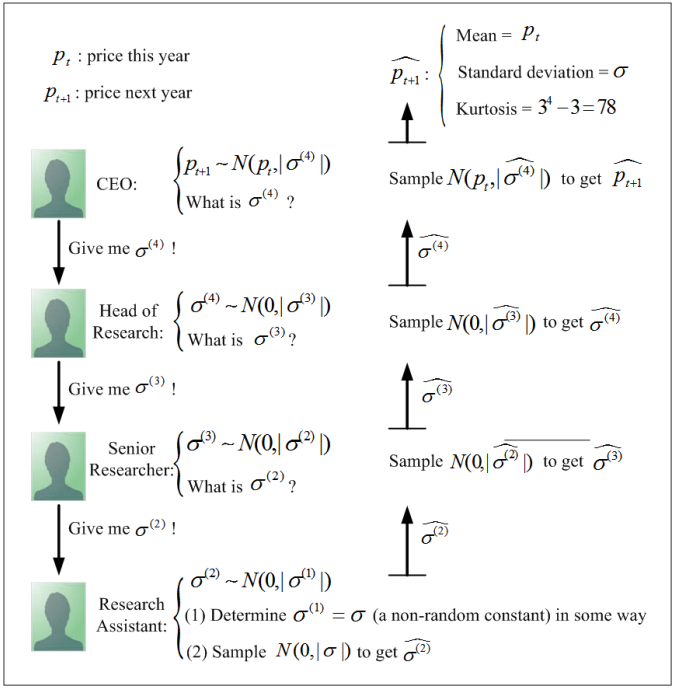

Fig. 1: A schematic example of using Gaussian-Chain distribution to model hierarchical structure.

It is obvious that the structure schematized in Fig. 1 is very common in real life, therefore it is important to model the situation mathematically so that we can understand the situation more precisely and study the problems in details. We will define Gaussian-Chain distribution to model the situation in Fig. 1 and prove some basic properties of the Gaussian-Chain distribution to get some insights for the structures in Fig. 1. For example, we will prove that the mean, the standard deviation and the kurtosis of the CEO's estimates are $p_t$, $\sigma$ and $3^4 - 3 = 78$ (which is much larger than the Gaussian kurtosis zero), respectively, i.e. the distribution of the CEO's estimates is very heavy-tailed. Consequently, although each individual in Fig. 1 uses a Gaussian model, their hierarchical connection results in a distribution that is much more heavy-tailed than the Gaussian distribution.

In order to extract useful information from the noisy price data, we need to filter out the noise that is heavy-tail distributed. The conventional least-squares algorithms are very sensitive to heavy-tailed disturbances [22], it is therefore important to construct filters that take the heavy-tailed distribution explicitly into consideration. We will develop two maximum likelihood type of filters for noises with Gaussian-Chain distributions, and show through simulations that these Gaussian-Chain filters are much better than the ordinary least-squares algorithm (which is optimal in the Gaussian framework) when the noises are heavy-tail distributed. We will use the Gaussian-Chain filters to extract useful information from real stock prices to detect the presence of big buyers and big sellers in the market.

This paper is organized as follows. In Section II, we will define the Gaussian-Chain distributions and prove their basic properties. In Sections III and IV, we will derive two filters --- 2$^{nd}$ and 3$^{rd}$ order Gaussian-Chain filters --- to filter out noises that are modeled as 2$^{nd}$ and 3$^{rd}$ order Gaussian-Chain distributions, respectively. In Section V, we will propose a model for big buyers and big sellers in stock market and use the Gaussian-Chain filters to estimate the parameters in the model; based on these parameter estimates we will develop a trading strategy (named Ride-the-Mood) and test it for some Hong Kong stocks. Section VI concludes the paper.

## II. THE GAUSSIAN-CHAIN DISTRIBUTION

First, we define Gaussian-Chain random variable as follows.

**Definition 1:** Let $N(m, \sigma)$ denote Gaussian distribution with mean $m$ and standard deviation $\sigma$. The *q'th-order Gaussian-Chain (GC) with parameters m and $\sigma$*, denoted as $\varepsilon_{m,\sigma}^{(q)}$, is a random variable:

$$\varepsilon_{m,\sigma}^{(q)} \sim N(m, |\sigma^{(q)}|) \qquad (1)$$

where $m$ is a constant and $\sigma^{(q)}$ is a random variable defined recursively through:

$$\sigma^{(j)} \sim N(0, |\sigma^{(j-1)}|) \qquad (2)$$

where $j=q$, $q-1$, ..., 2 with $\sigma^{(1)} = \sigma$ being a non-random positive constant. If $m = 0$ and $\sigma = 1$, then the $\varepsilon_{m,\sigma}^{(q)} = \varepsilon_{0,1}^{(q)}$ is called *the standard q'th-order Gaussian-Chain*.

The density function of the *q*'th-order Gaussian-Chain is given in the following lemma.

**Lemma 1:** The density function of the *q*'th-order Gaussian-Chain $\varepsilon_{m,\sigma}^{(q)}$, denoted as $f_{\varepsilon_{m,\sigma}^{(q)}}(x)$, is given as follows:

$$f_{\varepsilon_{m,\sigma}^{(q)}}(x) = \int_{-\infty}^{\infty} \cdots \int_{-\infty}^{\infty} \frac{1}{|\sigma^{(q)}|\sqrt{2\pi}} e^{-\frac{1}{2}\left(\frac{x-m}{\sigma^{(q)}}\right)^2} \frac{1}{|\sigma^{(q-1)}|\sqrt{2\pi}} e^{-\frac{1}{2}\left(\frac{\sigma^{(q)}}{\sigma^{(q-1)}}\right)^2}$$
$$\cdots \frac{1}{|\sigma|\sqrt{2\pi}} e^{-\frac{1}{2}\left(\frac{\sigma^{(2)}}{\sigma}\right)^2} d\sigma^{(q)} \cdots d\sigma^{(2)} \qquad (3)$$

Proof: Applying the conditional density formula

$$f(x) = \int_{-\infty}^{\infty} f(x|y)f(y)dy \qquad (4)$$

repeatedly and noticing that

$$f_{\varepsilon_{m,\sigma}^{(q)}}(x|\sigma^{(q)}) = \frac{1}{|\sigma^{(q)}|\sqrt{2\pi}} e^{-\frac{1}{2}\left(\frac{x-m}{\sigma^{(q)}}\right)^2} \qquad (5)$$

according to (1) and

$$f_{\sigma^{(j)}}(\sigma^{(j)}|\sigma^{(j-1)}) = \frac{1}{|\sigma^{(j-1)}|\sqrt{2\pi}} e^{-\frac{1}{2}\left(\frac{\sigma^{(j)}}{\sigma^{(j-1)}}\right)^2} \qquad (6)$$

for $j=q, q-1, \ldots, 2$ according to (2) with $\sigma^{(1)} = \sigma$, we get

$$f_{\varepsilon_{m,\sigma}^{(q)}}(x) = \int_{-\infty}^{\infty} \cdots \int_{-\infty}^{\infty} f_{\varepsilon_{m,\sigma}^{(q)}}(x|\sigma^{(q)}) f_{\sigma^{(q)}}(\sigma^{(q)}|\sigma^{(q-1)})$$
$$\cdots f_{\sigma^{(2)}}(\sigma^{(2)}|\sigma^{(1)}) d\sigma^{(q)} \cdots d\sigma^{(2)}$$
$$= \int_{-\infty}^{\infty} \cdots \int_{-\infty}^{\infty} \frac{1}{|\sigma^{(q)}|\sqrt{2\pi}} e^{-\frac{1}{2}\left(\frac{x-m}{\sigma^{(q)}}\right)^2} \frac{1}{|\sigma^{(q-1)}|\sqrt{2\pi}} e^{-\frac{1}{2}\left(\frac{\sigma^{(q)}}{\sigma^{(q-1)}}\right)^2}$$
$$\cdots \frac{1}{|\sigma|\sqrt{2\pi}} e^{-\frac{1}{2}\left(\frac{\sigma^{(2)}}{\sigma}\right)^2} d\sigma^{(q)} \cdots d\sigma^{(2)} \qquad (7)$$

∎

The following lemma gives the basic statistics of the $q$'th-order Gaussian-Chain.

**Lemma 2:** The mean, variance, 3$^{rd}$ and 4$^{th}$ order center moments, and kurtosis of the $q$'th-order Gaussian-Chain $\varepsilon_{m,\sigma}^{(q)}$ are given as follows:

$$E\left\{\varepsilon_{m,\sigma}^{(q)}\right\} = m \qquad (8)$$

$$Var\left\{\varepsilon_{m,\sigma}^{(q)}\right\} = \sigma^2 \qquad (9)$$

$$E\left\{\left[\varepsilon_{m,\sigma}^{(q)} - E\left(\varepsilon_{m,\sigma}^{(q)}\right)\right]^3\right\} = 0 \qquad (10)$$

$$E\left\{\left[\varepsilon_{m,\sigma}^{(q)} - E\left(\varepsilon_{m,\sigma}^{(q)}\right)\right]^4\right\} = 3^q \sigma^4 \qquad (11)$$

$$kur\left(\varepsilon_{m,\sigma}^{(q)}\right) \equiv \frac{E\left\{\left[\varepsilon_{m,\sigma}^{(q)} - E\left(\varepsilon_{m,\sigma}^{(q)}\right)\right]^4\right\}}{\left(Var\left\{\varepsilon_{m,\sigma}^{(q)}\right\}\right)^2} - 3 = 3^q - 3 \qquad (12)$$

Proof: Substitute (3) into

$$E\left\{\varepsilon_{m,\sigma}^{(q)}\right\} \equiv \int_{-\infty}^{\infty} x f_{\varepsilon_{m,\sigma}^{(q)}}(x) dx$$

$$= \int_{-\infty}^{\infty} \cdots \int_{-\infty}^{\infty} \underbrace{\left[\int_{-\infty}^{\infty} \frac{x}{|\sigma^{(q)}|\sqrt{2\pi}} e^{-\frac{1}{2}\left(\frac{x-m}{\sigma^{(q)}}\right)^2} dx\right]}_{m} \frac{1}{|\sigma^{(q-1)}|\sqrt{2\pi}} e^{-\frac{1}{2}\left(\frac{\sigma^{(q)}}{\sigma^{(q-1)}}\right)^2}$$
$$\cdots \frac{1}{|\sigma^{(1)}|\sqrt{2\pi}} e^{-\frac{1}{2}\left(\frac{\sigma^{(2)}}{\sigma^{(1)}}\right)^2} d\sigma^{(q)} \cdots d\sigma^{(2)}$$

$$= m \int_{-\infty}^{\infty} \cdots \int_{-\infty}^{\infty} \underbrace{\left[\int_{-\infty}^{\infty} \frac{1}{|\sigma^{(q-1)}|\sqrt{2\pi}} e^{-\frac{1}{2}\left(\frac{\sigma^{(q)}}{\sigma^{(q-1)}}\right)^2} d\sigma^{(q)}\right]}_{1} \frac{1}{|\sigma^{(q-2)}|\sqrt{2\pi}} e^{-\frac{1}{2}\left(\frac{\sigma^{(q-1)}}{\sigma^{(q-2)}}\right)^2}$$
$$\cdots \frac{1}{|\sigma^{(1)}|\sqrt{2\pi}} e^{-\frac{1}{2}\left(\frac{\sigma^{(2)}}{\sigma^{(1)}}\right)^2} d\sigma^{(q-1)} \cdots d\sigma^{(2)}$$

$$= \cdots = m \qquad (13)$$

Similarly,

$$Var\left\{\varepsilon_{m,\sigma}^{(q)}\right\} \equiv \int_{-\infty}^{\infty} \left[x - E(\varepsilon_{m,\sigma}^{(q)})\right]^2 f_{\varepsilon_{m,\sigma}^{(q)}}(x) dx$$

$$= \int_{-\infty}^{\infty} \cdots \int_{-\infty}^{\infty} \underbrace{\left[\int_{-\infty}^{\infty} \frac{(x-m)^2}{|\sigma^{(q)}|\sqrt{2\pi}} e^{-\frac{1}{2}\left(\frac{x-m}{\sigma^{(q)}}\right)^2} dx\right]}_{\left(\sigma^{(q)}\right)^2} \frac{1}{|\sigma^{(q-1)}|\sqrt{2\pi}} e^{-\frac{1}{2}\left(\frac{\sigma^{(q)}}{\sigma^{(q-1)}}\right)^2}$$
$$\cdots \frac{1}{|\sigma^{(1)}|\sqrt{2\pi}} e^{-\frac{1}{2}\left(\frac{\sigma^{(2)}}{\sigma^{(1)}}\right)^2} d\sigma^{(q)} \cdots d\sigma^{(2)}$$

$$= \int_{-\infty}^{\infty} \cdots \int_{-\infty}^{\infty} \underbrace{\left[\int_{-\infty}^{\infty} \frac{\left(\sigma^{(q)}\right)^2}{|\sigma^{(q-1)}|\sqrt{2\pi}} e^{-\frac{1}{2}\left(\frac{\sigma^{(q)}}{\sigma^{(q-1)}}\right)^2} d\sigma^{(q)}\right]}_{\left(\sigma^{(q-1)}\right)^2} \frac{1}{|\sigma^{(q-2)}|\sqrt{2\pi}} e^{-\frac{1}{2}\left(\frac{\sigma^{(q-1)}}{\sigma^{(q-2)}}\right)^2}$$
$$\cdots \frac{1}{|\sigma^{(1)}|\sqrt{2\pi}} e^{-\frac{1}{2}\left(\frac{\sigma^{(2)}}{\sigma^{(1)}}\right)^2} d\sigma^{(q-1)} \cdots d\sigma^{(2)}$$

$$= \cdots = \left(\sigma^{(1)}\right)^2 = \sigma^2 \qquad (14)$$

and,

$$E\left\{\left[\varepsilon_{m,\sigma}^{(q)} - E\left(\varepsilon_{m,\sigma}^{(q)}\right)\right]^3\right\} = \int_{-\infty}^{\infty} \left[x - E(\varepsilon_{m,\sigma}^{(q)})\right]^3 f_{\varepsilon_{m,\sigma}^{(q)}}(x) dx$$

$$= \int_{-\infty}^{\infty} \cdots \int_{-\infty}^{\infty} \underbrace{\left[\int_{-\infty}^{\infty} \frac{(x-m)^3}{|\sigma^{(q)}|\sqrt{2\pi}} e^{-\frac{1}{2}\left(\frac{x-m}{\sigma^{(q)}}\right)^2} dx\right]}_{0} \frac{1}{|\sigma^{(q-1)}|\sqrt{2\pi}} e^{-\frac{1}{2}\left(\frac{\sigma^{(q)}}{\sigma^{(q-1)}}\right)^2}$$
$$\cdots \frac{1}{|\sigma^{(1)}|\sqrt{2\pi}} e^{-\frac{1}{2}\left(\frac{\sigma^{(2)}}{\sigma^{(1)}}\right)^2} d\sigma^{(q)} \cdots d\sigma^{(2)}$$

$$= 0 \qquad (15)$$

Finally,

$$E\left\{\left[\varepsilon_{m,\sigma}^{(q)} - E\left(\varepsilon_{m,\sigma}^{(q)}\right)\right]^4\right\} = \int_{-\infty}^{\infty} \left[x - E(\varepsilon_{m,\sigma}^{(q)})\right]^4 f_{\varepsilon_{m,\sigma}^{(q)}}(x) dx$$

$$= \int_{-\infty}^{\infty} \cdots \int_{-\infty}^{\infty} \underbrace{\left[\int_{-\infty}^{\infty} \frac{(x-m)^4}{|\sigma^{(q)}|\sqrt{2\pi}} e^{-\frac{1}{2}\left(\frac{x-m}{\sigma^{(q)}}\right)^2} dx\right]}_{3\left(\sigma^{(q)}\right)^4} \frac{1}{|\sigma^{(q-1)}|\sqrt{2\pi}} e^{-\frac{1}{2}\left(\frac{\sigma^{(q)}}{\sigma^{(q-1)}}\right)^2}$$
$$\cdots \frac{1}{|\sigma^{(1)}|\sqrt{2\pi}} e^{-\frac{1}{2}\left(\frac{\sigma^{(2)}}{\sigma^{(1)}}\right)^2} d\sigma^{(q)} \cdots d\sigma^{(2)}$$

$$= 3 \int_{-\infty}^{\infty} \cdots \int_{-\infty}^{\infty} \underbrace{\left[\int_{-\infty}^{\infty} \frac{\left(\sigma^{(q)}\right)^4}{|\sigma^{(q-1)}|\sqrt{2\pi}} e^{-\frac{1}{2}\left(\frac{\sigma^{(q)}}{\sigma^{(q-1)}}\right)^2} d\sigma^{(q)}\right]}_{3\left(\sigma^{(q-1)}\right)^4} \frac{1}{|\sigma^{(q-2)}|\sqrt{2\pi}} e^{-\frac{1}{2}\left(\frac{\sigma^{(q-1)}}{\sigma^{(q-2)}}\right)^2}$$
$$\cdots \frac{1}{|\sigma^{(1)}|\sqrt{2\pi}} e^{-\frac{1}{2}\left(\frac{\sigma^{(2)}}{\sigma^{(1)}}\right)^2} d\sigma^{(q-1)} \cdots d\sigma^{(2)}$$

$$= \cdots = 3^q \left(\sigma^{(1)}\right)^4 = 3^q \sigma^4 \qquad (16)$$

∎

From (8), (9) and (12) we see that the mean and the standard deviation of the $q$'th-order Gaussian-Chain $\varepsilon_{m,\sigma}^{(q)}$ are equal to $m$ and $\sigma$, respectively, which are independent of order $q$ of the Gaussian-Chain, but the kurtosis of $\varepsilon_{m,\sigma}^{(q)}$ equals $3^q - 3$ which increases exponentially with order $q$ and is independent of the parameters $m$ and $\sigma$; that is, the Gaussian-Chain $\varepsilon_{m,\sigma}^{(q)}$ ($q>1$) and the Gaussian random variable $\varepsilon_{m,\sigma}^{(1)} \sim N(m,\sigma)$ have the same mean and variance, but the Gaussian-Chain kurtosis $3^q - 3$ is much larger the Gaussian kurtosis $3^1 - 3 = 0$, meaning that the Gaussian-Chain has a heavy-tailed distribution.

Fig. 2 plots the distribution functions of the standard Gaussian-Chain $\varepsilon_{0,1}^{(q)}$:

$$F_{\varepsilon_{0,1}^{(q)}}(x) \equiv \int_x^\infty f_{\varepsilon_{0,1}^{(q)}}(y) dy \quad (17)$$

for $q=1$ (standard Gaussian), 5, 10 and 20 in the log-log scale, and Table 1 gives the values of $F_{\varepsilon_{0,1}^{(q)}}(x)$ from $q=1$ to $q=20$ for $x=1,2,...,9$ (one to nine times of the standard deviation), which were obtained through Monte-Carlo simulations of $N=7,500,000$ samples. From Fig. 1 we see that the tails of the distributions of $\varepsilon_{0,1}^{(q)}$ are getting flatter and flatter as $q$ increases, verifying the heavy-tailed property of the Gaussian-Chain distributions, and the numbers in Table 1 show exactly how heavy the tails of the Gaussian-Chains are.

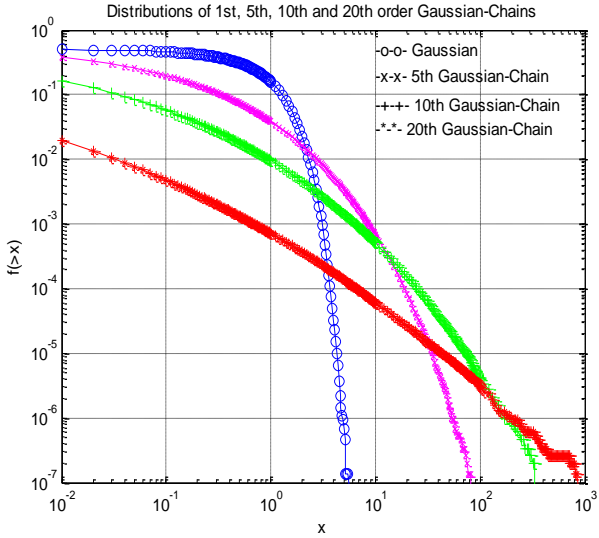

Fig. 2: The distribution functions $F_{\varepsilon_{0,1}^{(q)}}(x)$ for $q=1$ (Gaussian), 5, 10 and 20.

Similar to the relationship between Gaussian and standard Gaussian random variables, the following lemma shows that a Gaussian-Chain random variable can be easily converted to a standard Gaussian-Chain variable.

**Lemma 3:** The $\varepsilon_{m,\sigma}^{(q)}$ and $\varepsilon_{0,1}^{(q)}$ are convertible to each other through the following equation:

$$\varepsilon_{0,1}^{(q)} = \frac{\varepsilon_{m,\sigma}^{(q)} - m}{\sigma} \quad (18)$$

Proof: Let $\varepsilon = \frac{\varepsilon_{m,\sigma}^{(q)} - m}{\sigma}$ and its density function is

$$f_\varepsilon(x) = \frac{f_{\varepsilon_{m,\sigma}^{(q)}}(m + \sigma x)}{\left|\frac{\partial \varepsilon}{\partial \varepsilon_{m,\sigma}^{(q)}}\right|} = \sigma f_{\varepsilon_{m,\sigma}^{(q)}}(m + \sigma x) \quad (19)$$

Substituting the density function (3) into (19) and changing variables $w^{(j)} = \sigma^{(j)}/\sigma$ ($j=q,q-1,...,2$) yield

$$f_\varepsilon(x) = \int_{-\infty}^{\infty} \cdots \int_{-\infty}^{\infty} \frac{1}{|\sigma^{(q)}/\sigma|\sqrt{2\pi}} e^{-\frac{1}{2}\left(\frac{x}{\sigma^{(q)}/\sigma}\right)^2} \frac{1}{|\sigma^{(q-1)}|\sqrt{2\pi}} e^{-\frac{1}{2}\left(\frac{\sigma^{(q)}}{\sigma^{(q-1)}}\right)^2}$$
$$\cdots \frac{1}{|\sigma|\sqrt{2\pi}} e^{-\frac{1}{2}\left(\frac{\sigma^{(2)}}{\sigma}\right)^2} d\sigma^{(q)} \cdots d\sigma^{(2)}$$

$$= \int_{-\infty}^{\infty} \cdots \int_{-\infty}^{\infty} \frac{1}{|w^{(q)}|\sqrt{2\pi}} e^{-\frac{1}{2}\left(\frac{x}{w^{(q)}}\right)^2} \frac{1}{|\sigma^{(q-1)}/\sigma|\sqrt{2\pi}} e^{-\frac{1}{2}\left(\frac{w^{(q)}}{\sigma^{(q-1)}/\sigma}\right)^2}$$
$$\cdots \frac{1}{|\sigma|\sqrt{2\pi}} e^{-\frac{1}{2}\left(\frac{\sigma^{(2)}}{\sigma}\right)^2} dw^{(q)} d\sigma^{(q-1)} \cdots d\sigma^{(2)}$$

$$= \cdots$$

$$= \int_{-\infty}^{\infty} \cdots \int_{-\infty}^{\infty} \frac{1}{|w^{(q)}|\sqrt{2\pi}} e^{-\frac{1}{2}\left(\frac{x}{w^{(q)}}\right)^2} \frac{1}{|w^{(q-1)}|\sqrt{2\pi}} e^{-\frac{1}{2}\left(\frac{w^{(q)}}{w^{(q-1)}}\right)^2}$$
$$\cdots \frac{1}{|\sigma/\sigma|\sqrt{2\pi}} e^{-\frac{1}{2}\left(\frac{w^{(2)}}{\sigma/\sigma}\right)^2} dw^{(q)} \cdots dw^{(2)} \quad (20)$$

which is the density function of the standard Gaussian-Chain $\varepsilon_{0,1}^{(q)}$. ∎

## III. THE $2^{\text{ND}}$-ORDER GAUSSIAN-CHAIN FILTER

Let $p_t$ ($t = 0,1,2,...$) be the price of an asset (such as a stock) and $r_t \equiv \ln(p) - \ln(p_{t-1}) \cong (p_t - p_{t-1})/p_{t-1}$ be the return. Consider the price dynamical model

$$r_t = a^T ed(x_{t-1}) + \sigma \varepsilon_t^{(q)} \quad (21)$$

where $x_{t-1}$ is some function of the past prices up to $p_{t-1}$, $ed(x_{t-1})$ is a $m$-dimensional vector of excess demands (demand minus supply) from traders using different trading strategies, $a \in R^m$ is unknown parameter vector representing the relative strengths of the traders, and $\varepsilon_t^{(q)} = \varepsilon_{0,1}^{(q)}$ is i.i.d. random process with standard $q$'th-order Gaussian-Chain distribution. We assume that $ed(x_{t-1})$ is given (representing our model of the traders; a detailed model will be proposed in Section 5) and for any $x_{t-1}$ and $i, j = 1, 2, ..., m$, $ed_i(x_{t-1})ed_j(x_{t-1}) = 0$ if $i \neq j$ and $ed_i(x_{t-1})ed_j(x_{t-1}) \neq 0$ if $i = j$ (meaning that the traders are "orthogonal" to each other; non-orthogonal traders can be made orthogonal through the standard Gram-Schmidt scheme [28]). Our task is to estimate the parameter vector $a$ based on the information set $\{(r_j, x_{j-1}); j = 1, 2, ..., t\}$ up to the current time t.

Table 1: Values of $2F_{\varepsilon_{0,1}^{(q)}}(x)$ for different q and x.

| q | $1 - 2F_{\varepsilon_{0,1}^{(q)}}(1)$ (in %) | $2F_{\varepsilon_{0,1}^{(q)}}(x)$ (in %) | | | | | | | |
|---|---|---|---|---|---|---|---|---|---|
| | | x = 2 | x = 3 | x = 4 | x = 5 | x = 6 | x = 7 | x = 8 | x = 9 |
| 1 | 68.2805 | 4.5430 | 0.2716 | 0.0063 | 0.0001 | 0 | 0 | 0 | 0 |
| 2 | 79.0911 | 6.2040 | 1.9751 | 0.6521 | 0.2174 | 0.0745 | 0.0257 | 0.0089 | 0.0032 |
| 3 | 85.3445 | 5.3019 | 2.3084 | 1.1163 | 0.5801 | 0.3138 | 0.1750 | 0.1022 | 0.0604 |
| 4 | 89.4159 | 4.2239 | 2.1083 | 1.1731 | 0.7051 | 0.4439 | 0.2925 | 0.1977 | 0.1367 |
| 5 | 92.2286 | 3.2868 | 1.7593 | 1.0627 | 0.6893 | 0.4712 | 0.3338 | 0.2423 | 0.1788 |
| 6 | 94.2043 | 2.5493 | 1.4336 | 0.9097 | 0.6216 | 0.4419 | 0.3254 | 0.2473 | 0.1919 |
| 7 | 95.6433 | 1.9706 | 1.1459 | 0.7509 | 0.5263 | 0.3860 | 0.2943 | 0.2317 | 0.1846 |
| 8 | 96.7083 | 1.5144 | 0.9032 | 0.6061 | 0.4368 | 0.3289 | 0.2555 | 0.2019 | 0.1619 |
| 9 | 97.5143 | 1.1651 | 0.7057 | 0.4786 | 0.3481 | 0.2663 | 0.2102 | 0.1701 | 0.1401 |
| 10 | 98.1017 | 0.9068 | 0.5580 | 0.3842 | 0.2851 | 0.2219 | 0.1768 | 0.1449 | 0.1214 |
| 11 | 98.5463 | 0.7018 | 0.4367 | 0.3042 | 0.2273 | 0.1784 | 0.1441 | 0.1189 | 0.0995 |
| 12 | 98.8814 | 0.5444 | 0.3421 | 0.2403 | 0.1828 | 0.1432 | 0.1160 | 0.0960 | 0.0820 |
| 13 | 99.1380 | 0.4260 | 0.2694 | 0.1922 | 0.1454 | 0.1150 | 0.0942 | 0.0787 | 0.0665 |
| 14 | 99.3399 | 0.3279 | 0.2094 | 0.1501 | 0.1138 | 0.0906 | 0.0739 | 0.0623 | 0.0531 |
| 15 | 99.4932 | 0.2532 | 0.1629 | 0.1174 | 0.0899 | 0.0715 | 0.0588 | 0.0490 | 0.0418 |
| 16 | 99.6060 | 0.1956 | 0.1271 | 0.0919 | 0.0705 | 0.0574 | 0.0478 | 0.0404 | 0.0350 |
| 17 | 99.6944 | 0.1550 | 0.1013 | 0.0731 | 0.0567 | 0.0458 | 0.0377 | 0.0318 | 0.0275 |
| 18 | 99.7652 | 0.1179 | 0.0770 | 0.0561 | 0.0437 | 0.0352 | 0.0294 | 0.0250 | 0.0218 |
| 19 | 99.8194 | 0.0906 | 0.0590 | 0.0431 | 0.0338 | 0.0272 | 0.0230 | 0.0194 | 0.0168 |
| 20 | 99.8565 | 0.0739 | 0.0491 | 0.0359 | 0.0278 | 0.0225 | 0.0188 | 0.0160 | 0.0137 |

In this section we consider the $q=2$ case. The following algorithm gives the maximum likelihood solution to the problem.

**2nd-order Gaussian-Chain (GC) Filter:** Consider model (21) with $q=2$. Given $\{(r_j, x_{j-1}); j = 1, 2, ..., t\}$, the estimate of the parameter vector $a$, denoted by $\hat{a}_t = (\hat{a}_{t,1}, \hat{a}_{t,2}, ..., \hat{a}_{t,m})^T$, can be computed through the following recursive algorithm:

$$\hat{a}_{t,i} = \left(\frac{\lambda v_t^2 b_{t-1,i}}{\lambda v_t^2 b_{t-1,i} + ed_i^2(x_{t-1})}\right)\hat{a}_{t-1,i} + \frac{r_t\, ed_i(x_{t-1})}{\lambda v_t^2 b_{t-1,i} + ed_i^2(x_{t-1})} \quad (22)$$

$$b_{t,i} = \lambda\, b_{t-1,i} + \frac{ed_i^2(x_{t-1})}{v_t^2} \quad (23)$$

$$v_t^2 = \frac{\sqrt{\hat{\sigma}_{t-1}^4 + 4\hat{\sigma}_{t-1}^2\left(r_t - \hat{a}_{t-1}^T ed(x_{t-1})\right)^2} - \hat{\sigma}_{t-1}^2}{2} \quad (24)$$

with

$$\hat{\sigma}_t^2 = \left(\frac{\lambda c_{t-1}}{1 + \lambda c_{t-1}}\right)\hat{\sigma}_{t-1}^2 + \frac{v_t^2}{1 + \lambda c_{t-1}} \quad (25)$$

$$c_t = 1 + \lambda c_{t-1} \quad (26)$$

where $t = 1, 2, 3, ...$, $i = 1, 2, ..., m$, $\lambda \in (0,1)$ is a weighting factor, and initial condition $\hat{a}_{0,i} = b_{0,i} = c_0 = 0$ and $\hat{\sigma}_0 = 1$.

Deviation of the 2$^{nd}$-order Gaussian-Chain Filter: Substituting $\sigma\varepsilon_j^{(2)} = r_j - a^T ed(x_{j-1})$ into the joint density function of $\sigma\varepsilon_j^{(2)} = \sigma\varepsilon_{0,1}^{(2)} = \varepsilon_{0,\sigma}^{(2)}$ ($j = 1,2,\ldots,t$), we get

$$f(r_1, x_0, \ldots, r_t, x_{t-1}) =$$
$$\int_{-\infty}^{\infty}\cdots\int_{-\infty}^{\infty}\prod_{j=1}^{t}\left(\frac{1}{|v_j|\sqrt{2\pi}}e^{-\frac{1}{2}\left(\frac{r_j-a^T ed(x_{j-1})}{v_j}\right)^2}\frac{1}{\sigma\sqrt{2\pi}}e^{-\frac{1}{2}\left(\frac{v_j}{\sigma}\right)^2}\right)dv_1\cdots dv_t \quad (27)$$

which is the integration of the $t$-dimensional function

$$g(v_1,\ldots,v_t) = \prod_{j=1}^{t}\left(\frac{1}{|v_j|\sqrt{2\pi}}e^{-\frac{1}{2}\left(\frac{r_j-a^T ed(x_{j-1})}{v_j}\right)^2}\frac{1}{\sigma\sqrt{2\pi}}e^{-\frac{1}{2}\left(\frac{v_j}{\sigma}\right)^2}\right) \quad (28)$$

over $R^t$. The traditional maximum likelihood approach is to find the parameters $a$ and $\sigma$ such that the $f(r_1, x_0, \ldots, r_t, x_{t-1})$ of (27) is maximized, but this will not give simple solutions due to the integrations in (27). So here we take a different approach by viewing $v_1, \ldots, v_t$ as free parameters and find $a$, $\sigma$, together with $v_1, \ldots, v_t$, to maximize $g(v_1, \ldots, v_t)$ of (28). Conceptually, the traditional approach is to find $a$ and $\sigma$ to maximize the integration of the function $g(v_1, \ldots, v_t)$ over $R^t$, whereas our approach is to find $a$ and $\sigma$ to maximize the maximum of $g(v_1, \ldots, v_t)$ over $R^t$. Given $\{(r_j, x_{j-1}); j = 1,2,\ldots,t\}$ and considering the more general case that the parameters $a$ are slow time-varying, we give more weights to more recent data and define the likelihood function:

$$L(a, \sigma, v_1, \ldots, v_t) = \prod_{j=1}^{t}\left(\frac{1}{|v_j|\sqrt{2\pi}}e^{-\frac{1}{2}\left(\frac{r_j-a^T ed(x_{j-1})}{v_j}\right)^2}\frac{1}{\sigma\sqrt{2\pi}}e^{-\frac{1}{2}\left(\frac{v_j}{\sigma}\right)^2}\right)^{\lambda^{t-j}}$$

$$= \frac{e^{-\frac{1}{2}\sum_{j=1}^{t}\left[\left(\frac{r_j-a^T ed(x_{j-1})}{v_j}\right)^2+\left(\frac{v_j}{\sigma}\right)^2\right]\lambda^{t-j}}}{\prod_{j=1}^{t}(|v_j|2\pi\sigma)^{\lambda^{t-j}}} \quad (29)$$

where $\lambda \in (0,1)$ is a weighting factor. The weighting scheme in (29) makes the likelihood function more sensitive to recent data when the parameters change. Taking the partial derivative of the likelihood function (29) with respect to the parameter vector $a$ and setting it to zero:

$$\frac{\partial L(a,\sigma,v_1,\ldots,v_t)}{\partial a} =$$
$$L(a,\sigma,v_1,\ldots,v_t)\sum_{j=1}^{t}\left(\frac{r_j-a^T ed(x_{j-1})}{v_j^2}\right)\lambda^{t-j}ed(x_{j-1}) = 0 \quad (30)$$

we get

$$\sum_{j=1}^{t}\left(\frac{r_j-\hat{a}_t^T ed(x_{j-1})}{v_j^2}\right)\lambda^{t-j}ed(x_{j-1}) = 0 \quad (31)$$

where $\hat{a}_t$ denotes the maximum likelihood estimate of the parameter vector $a$. From (31) and using the orthogonal and non-vanishing assumption for $ed(x_{j-1})$, we obtain

$$\hat{a}_t = \left[\sum_{j=1}^{t}\left(\frac{ed(x_{j-1})ed^T(x_{j-1})}{v_j^2/\lambda^{t-j}}\right)\right]^{-1}\sum_{j=1}^{t}\left(\frac{r_j}{v_j^2/\lambda^{t-j}}\right)ed(x_{j-1})$$

$$= \begin{pmatrix} \dfrac{\sum_{j=1}^{t}\left(\dfrac{r_j}{\frac{v_j^2}{\lambda^{t-j}}}\right)ed_1(x_{j-1})}{\sum_{j=1}^{t}\left(\dfrac{ed_1^2(x_{j-1})}{\frac{v_j^2}{\lambda^{t-j}}}\right)} \\ \vdots \\ \dfrac{\sum_{j=1}^{t}\left(\dfrac{r_j}{\frac{v_j^2}{\lambda^{t-j}}}\right)ed_m(x_{j-1})}{\sum_{j=1}^{t}\left(\dfrac{ed_m^2(x_{j-1})}{\frac{v_j^2}{\lambda^{t-j}}}\right)} \end{pmatrix} \quad (32)$$

or

$$\hat{a}_{t,i} = \frac{\sum_{j=1}^{t}\left(\frac{r_j}{v_j^2/\lambda^{t-j}}\right)ed_i(x_{j-1})}{\sum_{j=1}^{t}\left(\frac{ed_i^2(x_{j-1})}{v_j^2/\lambda^{t-j}}\right)} \quad (33)$$

$i = 1,2,\ldots,m$. To get a recursive formula to compute the $\hat{a}_{t,i}$ of (33), define

$$b_{t,i} \equiv \sum_{j=1}^{t}\left(\frac{ed_i^2(x_{j-1})}{v_j^2/\lambda^{t-j}}\right)$$
$$= \frac{ed_i^2(x_{t-1})}{v_t^2} + \sum_{j=1}^{t-1}\left(\frac{ed_i^2(x_{j-1})}{v_j^2/\lambda^{t-1-j}}\right)\lambda$$
$$= \lambda\, b_{t-1,i} + \frac{ed_i^2(x_{t-1})}{v_t^2} \quad (34)$$

and rewrite (33) as

$$\left(\lambda\, b_{t-1,i} + \frac{ed_i^2(x_{t-1})}{v_t^2}\right)\hat{a}_{t,i} =$$
$$\left(\frac{r_t}{v_t^2}\right)ed_i(x_{t-1}) + \lambda\sum_{j=1}^{t-1}\left(\frac{r_j}{v_j^2/\lambda^{t-1-j}}\right)ed_i(x_{j-1}) \quad (35)$$

Dividing both sides of (35) by $b_{t-1,i}$ and noticing (33), we have

$$\left(\lambda + \frac{ed_i^2(x_{t-1})}{v_t^2 b_{t-1,i}}\right)\hat{a}_{t,i} = \left(\frac{r_t}{v_t^2 b_{t-1,i}}\right)ed_i(x_{t-1}) + \lambda\,\hat{a}_{t-1,i} \quad (36)$$

or

$$\hat{a}_{t,i} = \left(\frac{\lambda b_{t-1,i}v_t^2}{\lambda v_t^2 b_{t-1,i} + ed_i^2(x_{t-1})}\right)\hat{a}_{t-1,i}$$
$$+ \frac{r_t\, ed_i(x_{t-1})}{\lambda v_t^2 b_{t-1,i} + ed_i^2(x_{t-1})} \quad (37)$$

(37) and (34) are (22) and (23), respectively.

We now proceed to find the maximum likelihood estimate of $\sigma$. Taking the partial derivative of the likelihood function (29) with respect to $\sigma$ and setting it to zero:

$$\frac{\partial L(a, \sigma, v_1, \ldots, v_t)}{\partial \sigma} =$$

$$L(a, \sigma, v_1, \ldots, v_t) \left( \frac{\sum_{j=1}^{t} v_j^2 \lambda^{t-j}}{\sigma^3} - \frac{\sum_{j=1}^{t} \lambda^{t-j}}{\sigma} \right) = 0 \quad (38)$$

we get the estimate of $\sigma$, denoted by $\hat{\sigma}_t$, as

$$\hat{\sigma}_t = \sqrt{\frac{\sum_{j=1}^{t} v_j^2 \lambda^{t-j}}{\sum_{j=1}^{t} \lambda^{t-j}}} \quad (39)$$

Defining

$$c_t \equiv \sum_{j=1}^{t} \lambda^{t-j} = 1 + \lambda c_{t-1} \quad (40)$$

and reorganizing (39) we get the following recursive formula for computing $\hat{\sigma}_t$:

$$\hat{\sigma}_t^2 = \left( \frac{\lambda c_{t-1}}{1 + \lambda c_{t-1}} \right) \hat{\sigma}_{t-1}^2 + \frac{v_t^2}{1 + \lambda c_{t-1}} \quad (41)$$

which is (25), and (40) is (26).

Finally, we find the maximum likelihood estimate of $v_t$. Rewriting the likelihood function (29) as

$$L(a, \sigma, v_1, \ldots, v_t) =$$

$$[L(a, \sigma, v_1, \ldots, v_{t-1})]^\lambda \left( \frac{1}{|v_t| 2\pi\sigma} e^{-\frac{1}{2}\left[ \left( \frac{r_t - a^T ed(x_{t-1})}{v_t} \right)^2 + \left( \frac{v_t}{\sigma} \right)^2 \right]} \right) \quad (42)$$

we see that finding $v_t$ to maximize $L(a, \sigma, v_1, \ldots, v_t)$ is equivalent to maximizing

$$\bar{P}(r_t, x_{t-1}|a, \sigma, v_t) \equiv \frac{1}{|v_t| 2\pi\sigma} e^{-\frac{1}{2}\left[ \left( \frac{r_t - a^T ed(x_{t-1})}{v_t} \right)^2 + \left( \frac{v_t}{\sigma} \right)^2 \right]} \quad (43)$$

Taking the partial derivative of $\bar{P}(r_t, x_{t-1}|a, \sigma, v_t)$ with respect to $|v_t|$ and setting it to zero:

$$\frac{\partial \bar{P}(r_t, x_{t-1}|a, \sigma, v_t)}{\partial |v_t|} =$$

$$\bar{P}(r_t, x_{t-1}|a, \sigma, v_t) \left[ \frac{(r_t - a^T ed(x_{t-1}))^2}{|v_t|^3} - \frac{|v_t|}{\sigma^2} - \frac{1}{|v_t|} \right] = 0 \quad (44)$$

we get

$$\frac{|v_t|^4}{\sigma^2} + |v_t|^2 - (r_t - a^T ed(x_{t-1}))^2 = 0 \quad (45)$$

Since when estimating $v_t$ (at time t) the best available estimates of $a$ and $\sigma$ are $\hat{a}_{t-1}$ and $\hat{\sigma}_{t-1}$, respectively, we substitute the $a$ and $\sigma$ in (45) by $\hat{a}_{t-1}$ and $\hat{\sigma}_{t-1}$, respectively, and solve (45) to get

$$v_t^2 = \frac{\sqrt{\hat{\sigma}_{t-1}^4 + 4\hat{\sigma}_{t-1}^2 (r_t - \hat{a}_{t-1}^T ed(x_{t-1}))^2} - \hat{\sigma}_{t-1}^2}{2} \quad (46)$$

which is (24). ∎

Before we apply the 2nd-order Gaussian-Chain filter to the real stock data in Section 5, we now perform simulation to get a basic feeling of the performance of the GC filter. We consider a simple tracking problem (estimating a time-varying parameter) with heavy-tailed noise and compare the 2nd-order GC filter with the standard least-squares algorithm. The purpose of this simulation is to show that the least-squares algorithm is very sensitive to heavy-tailed noise, whereas the GC filter, which takes the heavy-tail explicitly into consideration, does much better.

Specifically, consider the special case of (21) with $ed(x_{t-1}) \equiv 1, \sigma = 40, q = 5$ and $a$ being a time-varying variable: $a = a(t) = 20\left[1 + sin\left(\frac{8\pi t}{1000}\right)\right]$, i.e. the data $r_t$ ($t = 1,2,3,\ldots$) are generated by $r_t = 20\left[1 + sin\left(\frac{8\pi t}{1000}\right)\right] + 40\varepsilon_{0,1}^{(5)}$, and our task is to estimate the time-varying $a(t)$ based on the data set $\{r_1, r_2, \ldots, r_t\}$ at every time point $t = 1,2,3,\ldots$. Although the 2nd-order GC filter is designed for 2nd-order GC noise, we intentionally add a 5th-order GC noise $40\varepsilon_{0,1}^{(5)}$ to the signal $a(t) = 20\left[1 + sin\left(\frac{8\pi t}{1000}\right)\right]$ to test the filter in this very difficult situation. Fig. 3 shows the simulation results, where the top sub-figure is $r_t$, the middle sub-figure is the true $a(t)$ (green) and its estimate $\hat{a}_t$ using the standard recursive least-squares algorithm with exponential forgetting (page 53 of [3]), and the bottom sub-figure plots the true $a(t)$ (green) and its estimate $\hat{a}_t$ using the 2nd-order GC filter. We see from Fig. 3 that the least-squares algorithm is very sensitive to the large disturbances that occur quite frequently under the heavy-tailed noise, whereas the GC filter is very robust and gives accurate estimate for the time-varying variable.

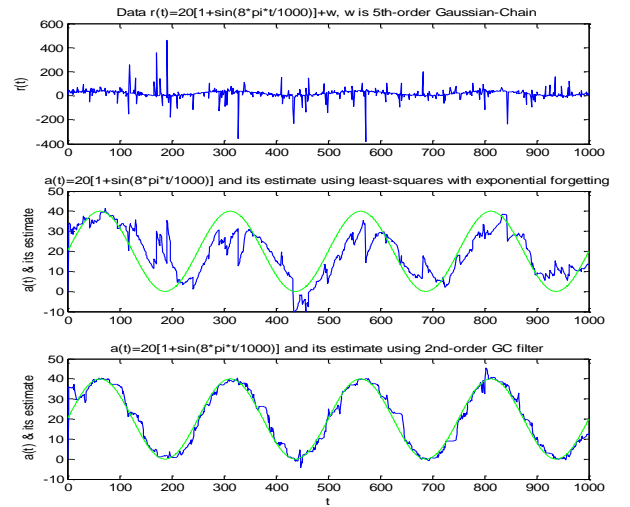

Fig. 3: Simulation results. Top: data $r_t$. Middle: true $a(t)$ (green) and its estimate $\hat{a}_t$ using the standard recursive least-squares algorithm with exponential forgetting. Bottom: true $a(t)$ (green) and its estimate $\hat{a}_t$ using the 2nd-order GC filter.

## IV. THE 3RD-ORDER GAUSSIAN-CHAIN FILTER

Similar to the 2nd-order GC filter developed in the last section, the following 3rd-order GC filter is designed to filter out 3rd-order Gaussian-Chain noise.

**3rd-order Gaussian-Chain (GC) Filter:** Consider model (21) with $q=3$. Given $\{(r_j, x_{j-1}); j = 1,2,...,t\}$, the estimate of the parameter vector $a$, denoted by $\hat{a}_t = (\hat{a}_{t,1}, \hat{a}_{t,2}, ..., \hat{a}_{t,m})^T$, can be computed through the following recursive algorithm:

$$\hat{a}_{t,i} = \left(\frac{\lambda v_t^2 b_{t-1,i}}{\lambda v_t^2 b_{t-1,i} + ed_i^2(x_{t-1})}\right)\hat{a}_{t-1,i} + \frac{r_t\, ed_i(x_{t-1})}{\lambda v_t^2 b_{t-1,i} + ed_i^2(x_{t-1})} \quad (47)$$

$$b_{t,i} = \lambda\, b_{t-1,i} + \frac{ed_i^2(x_{t-1})}{v_t^2} \quad (48)$$

$$v_t^2 = \frac{u_t^2(u_t^2 + \hat{\sigma}_{t-1}^2)}{\hat{\sigma}_{t-1}^2} \quad (49)$$

$$u_t^2 = \sqrt[3]{\frac{s}{2} + \sqrt{\left(\frac{s}{2}\right)^2 - \frac{\hat{\sigma}_{t-1}^{12}}{27}}} + \sqrt[3]{\frac{s}{2} - \sqrt{\left(\frac{s}{2}\right)^2 - \frac{\hat{\sigma}_{t-1}^{12}}{27}}} - \hat{\sigma}_{t-1}^2 \quad (50)$$

$$s = \hat{\sigma}_{t-1}^4\left[\left(r_t - \hat{a}_{t-1}^T ed(x_{t-1})\right)^2 + 4\hat{\sigma}_{t-1}^2\right] \quad (51)$$

with

$$\hat{\sigma}_t^2 = \left(\frac{\lambda c_{t-1}}{1 + \lambda c_{t-1}}\right)\hat{\sigma}_{t-1}^2 + \frac{u_t^2}{1 + \lambda c_{t-1}} \quad (52)$$

$$c_t = 1 + \lambda c_{t-1} \quad (53)$$

where $t = 1,2,3,...$, $i = 1,2,...,m$, $\lambda \in (0,1)$ is a weighting factor, and initial condition $\hat{a}_{0,i} = b_{0,i} = c_0 = 0$ and $\hat{\sigma}_0 = 1$.

Derivation of the 3rd-order Gaussian-Chain filter is given in the Appendix.

## V. DETECTING BIG BUYERS AND BIG SELLERS IN STOCK MARKET

Prices are determined by supply and demand. For stock market, the main supply and demand come from the institutional investors (pension funds, mutual funds, hedge funds, money managers, investment banks, etc.) who manage large sums of money and often buy or sell a stock in large quantity. So we make the following basic assumption:

**Assumption 1:** The stock prices are mainly determined by the actions of the institutional investors; we call them big buyers and big sellers.

The main problem facing the big buyers and the big sellers is that the amount of stocks available for sell or to buy at any time instant is very limited, so the big buyers and the big sellers have no choice but to buy or sell the stock incrementally (little by little) over a period of time lasting for weeks or even months [16]. Statistically, these actions of the big buyers and the big sellers cause the signed returns of the stocks being a long memory process --- the autocorrelation function of the signed returns decays very slowly --- a well-known stylized fact in finance ([6], [7]). Practically, these persistent buy or sell actions give the small investors a chance to detect the presence of the big buyers and the big sellers and follow them up ([29], [30]).

To make such detection, let's analyze in what situations the big buyers are most likely to buy and in what situations the big sellers are most likely to sell. Consider the case that the price is rising (the current price is above a moving average of the past prices), if the big buyers were still buying in this situation, the price would increase even further, resulting in high cost for the big buyers (buy the stock with higher prices). Similarly, if the big sellers were still selling when the price is declining, the price would decline even further, causing great increase of the cost for the big sellers (sell the stock with lower prices). Consequently, we have the following assumption:

**Assumption 2:** The big buyers (sellers) are buying (selling) if the price is decreasing (increasing); the larger the price decrease (increase), the stronger the buy (sell) actions from the big buyers (sellers); and the big buyers (sellers) take no action if the price is rising (declining).

Consider the price dynamical model (21). Based on Assumption 1, we choose $ed(x_{t-1}) = [ed_1(x_{t-1}), ed_2(x_{t-1})]^T$ and $a = [a_1, a_2]^T$, where $ed_1(x_{t-1})$ ($ed_2(x_{t-1})$) is the exceed demand from the big buyers (sellers) and $a_1$ ($a_2$) is the strength of the big buyers (sellers), and put the price influence from all other traders into the noise term $\sigma\varepsilon_t^{(q)}$. Define $x_t$ to be the log-ratio (relative change) of the price $p_t$ to its $n$-step moving average:

$$x_t = \ln\left(p_t \Big/ \frac{1}{n}\sum_{i=0}^{n-1} p_{t-i}\right) \quad (54)$$

so that $x_t > 0$ means the price is rising and $x_t < 0$ means the price is declining. Based on Assumption 2, we choose

$$ed_1(x_{t-1}) = \begin{cases} 0 & if\ x_{t-1} \geq 0 \\ |x_{t-1}| & if\ x_{t-1} < 0 \end{cases} \quad (55)$$

and

$$ed_2(x_{t-1}) = \begin{cases} 0 & if\ x_{t-1} \leq 0 \\ x_{t-1} & if\ x_{t-1} > 0 \end{cases} \quad (56)$$

This completes the specification of the price dynamical model (21).

Before we apply the GC filters with the price model (21) (with excess demand functions chosen as (55) and (56)) to real stock data, we perform some simulations to get a feeling of the performance of the GC filters for this very nonlinear model. Figs. 4 and 5 show the simulation results of the 2nd and 3rd-order GC filters, respectively, where the top sub-figures of Figs. 4 and 5 plot the true parameter $a_1(t)$ and $a_2(t)$ (piece-wise constants; blue lines) and their estimates $\hat{a}_{t,1}$ (red lines) and

$\hat{a}_{t,2}$ (green lines), and the bottom sub-figures plot the corresponding price series $p_t$'s generated by the price dynamical model (21) with parameters $\sigma = 0.01$ and $q = 2$ (3) for Fig. 4 (Fig. 5). The signal-to-noise ratio is defined as:

$$\frac{S}{N} = \frac{\left(\frac{1}{N}\sum_{t=1}^{N}\left(a_1(t)\,ed_1(x_{t-1}) + a_2(t)\,ed_2(x_{t-1})\right)^2\right)^{\frac{1}{2}}}{\left(\frac{1}{N}\sum_{t=1}^{N}\left(\sigma\varepsilon_t^{(q)}\right)^2\right)^{\frac{1}{2}}} \quad (57)$$

where N is the total number of data points. The signal-to-noise ratios of the price series in Figs. 4 and 5 are 1.0052 and 0.95429 (roughly equal to 1), respectively, meaning that the price impact of the big sellers plus the big buyers is roughly equal to the summation of the price impacts of the rest of the traders; and we see from Figs. 4 and 5 that in such situations the GC filters could estimate the parameters with some delay.

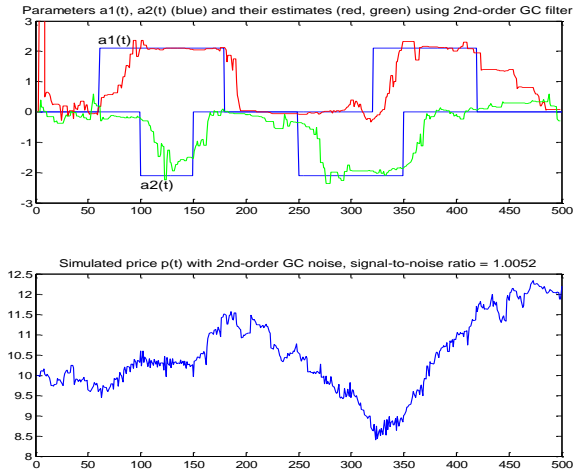

Fig. 4: Simulation of the 2nd-order GC filter. Top: true parameters (blue) and their estimates $\hat{a}_{t,1}$ (red) and $\hat{a}_{t,2}$ (green). Bottom: the price series $p_t$ (signal-to-noise ratio = 1.0052).

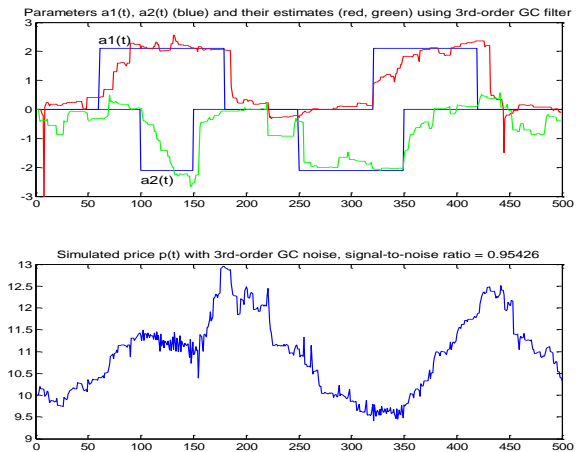

Fig. 5: Simulation of the 2nd-order GC filter. Top: true parameters (blue) and their estimates $\hat{a}_{t,1}$ (red) and $\hat{a}_{t,2}$ (green). Bottom: the price series $p_t$ (signal-to-noise ratio = 0.95429).

We now propose a trading strategy, called *Ride-the-Mood* (*RideMood*), based on the estimated strengths of the big buyers and the big sellers. Since positive $a_1$ (negative $a_2$) implies the presence of big buyer (big seller) and the absolute value of $a_1$ ($a_2$) represents the strength of the big buyer (big seller), the variable:

$$mood(t) \equiv \text{big buyer strength} - \text{big seller strength}$$
$$= \hat{a}_{t,1} + \hat{a}_{t,2} \quad (58)$$

can be viewed as the *mood* of the market: $mood(t) > 0$ implies the big buyers are in the upper hand over the big sellers, whereas $mood(t) < 0$ implies the opposite. The basic idea of the RideMood strategy is simply to buy when $mood(t)$ is changing from negative to positive and to sell when it is changing back from positive to negative. Fig. 6 gives the flow chart of the RideMood strategy, where we use the 5-day moving average of $mood(t)$:

$$\overline{mood}(t,5) \equiv \frac{1}{5}\sum_{i=1}^{5}(\hat{a}_{t-i+1,1} + \hat{a}_{t-i+1,2}) \quad (59)$$

to make the decision because $mood(t)$ itself is very noisy (see the simulations in Figs. 4 and 5) and using the moving average can reduce false signals.

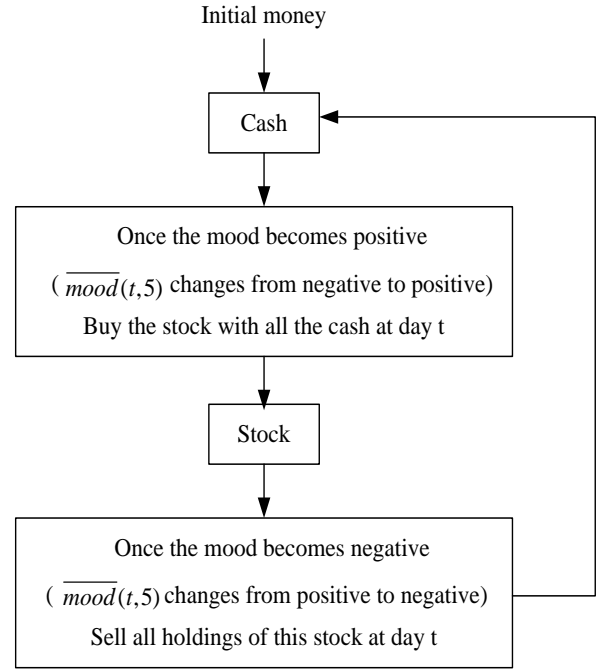

Fig. 6: The Ride-the-Mood (RideMood) trading strategy.

We now test the RideMood strategy for five blue-chip Hong Kong stocks: HK0005 (HSBC Holdings plc), HK0939 (China Construction Bank Corporation), HK0941 (China Mobile), HK1398 (Industrial and Commercial Bank of China), and HK3988 (Bank of China Ltd). We will use the daily closing prices of these stocks over the two-year period from April 2, 2012 to March 31, 2014 as the $p_t$ in the GC filters[1]. The results are shown in Figs. 7.1 to 7.5 and Table 2 for the 2$^{nd}$-order GC filter, and Figs. 8.1 to 8.5 and Table 3 for the 3$^{rd}$-order GC filter, where the top sub-figures in Figs. 7.1 to 7.5 (Figs. 8.1 to 8.5) plot the daily closing prices of the five stocks over the two-year period and the buy (green vertical lines) – sell (red vertical lines) points using the 2$^{nd}$-order (3$^{rd}$-order) GC filter in the RideMood scheme, the bottom sub-figures in Figs. 7.1 to 7.5 (Figs. 8.1 to 8.5) plot the $\overline{mood}(t,5)$ (59) using the 2$^{nd}$-order (3$^{rd}$-order) GC filter, and Table 2 (Table 3) gives the detailed buy/sell prices, dates and returns using the RideMood strategy with the two GC filters. The last rows in Tables 2 and 3 give the returns using the Buy-and-Hold strategy over the two-year period for the stocks (computed as $\frac{p(2014-03-31)-p(2012-04-02)}{p(2012-04-02)}$). From the last two rows in Tables 2 and 3 we can calculate that if we put these five stocks into a equally weighted portfolio, then over the two years from April 2, 2012 to March 31, 2014, the returns of the fortfolios using RideMood with 2$^{nd}$-order GC filter, RideMood with 3$^{rd}$-order GC filter, and Buy-and-Hold are 15.16%, 14.16%, and 5.56%, respectively. For reference the closing values of the Hang Seng Index were 22151 in March 31, 2014 and 20555 in April 2, 2012, representing a return of $\frac{22151-20555}{20555} = 7.76\%$. The good performance of the RideMood strategy with GC filters is apparent.

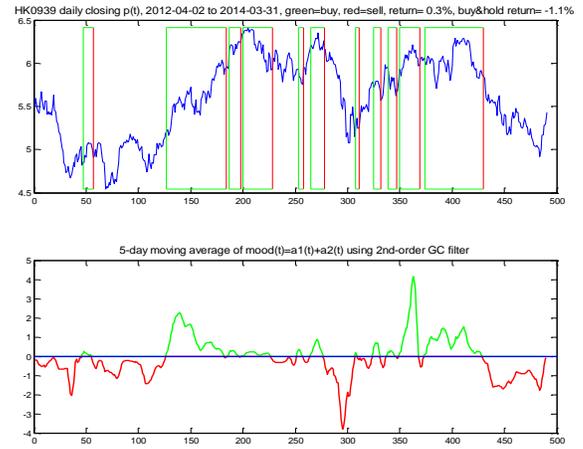

Fig. 7.2: Top: HK0939 daily closing $p_t$ from 2012-04-04 to 2014-03-31 and buy (green), sell (red) points using RideMood with 2$^{nd}$-order GC filter. Bottom: $\overline{mood}(t,5)$ (green when $\overline{mood}(t,5) > 0$ and red when $\overline{mood}(t,5) < 0$).

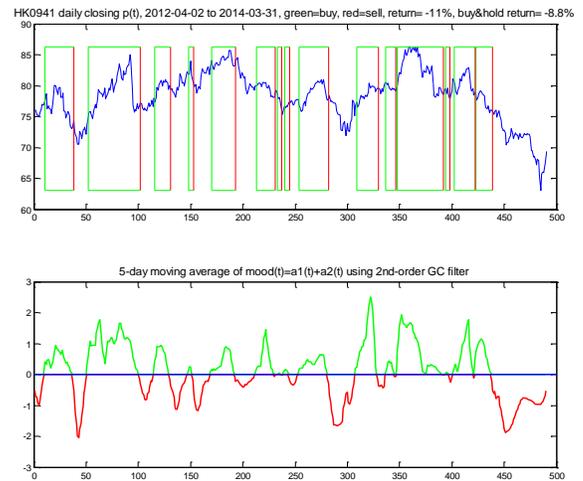

Fig. 7.3: Top: HK0941 daily closing $p_t$ from 2012-04-04 to 2014-03-31 and buy (green), sell (red) points using RideMood with 2$^{nd}$-order GC filter. Bottom: $\overline{mood}(t,5)$ (green when $\overline{mood}(t,5) > 0$ and red when $\overline{mood}(t,5) < 0$).

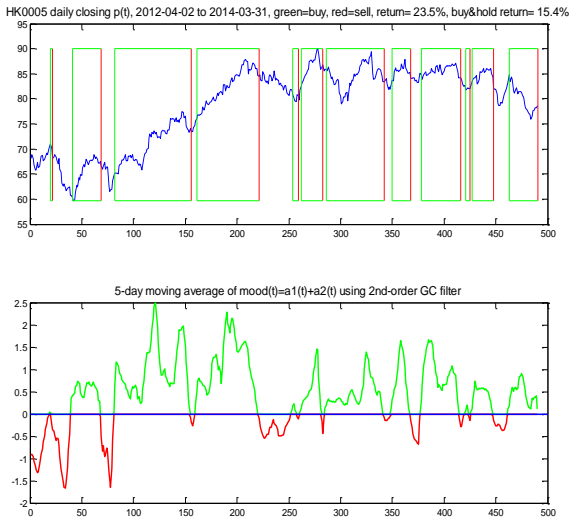

Fig. 7.1: Top: HK0005 daily closing $p_t$ from 2012-04-04 to 2014-03-31 and buy (green), sell (red) points using RideMood with 2$^{nd}$-order GC filter. Bottom: $\overline{mood}(t,5)$ (green when $\overline{mood}(t,5) > 0$ and red when $\overline{mood}(t,5) < 0$).

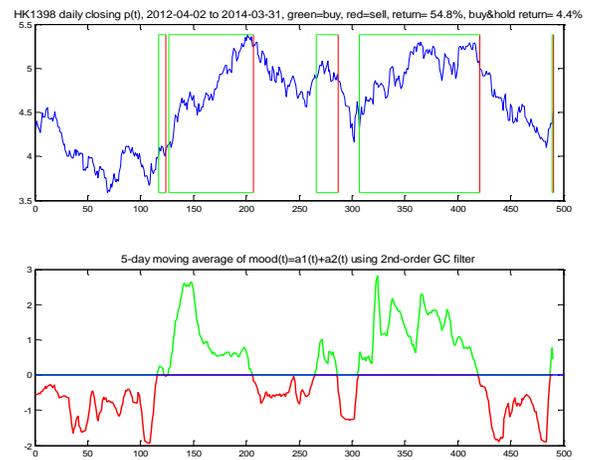

Fig. 7.4: Top: HK1398 daily closing $p_t$ from 2012-04-04 to 2014-03-31 and buy (green), sell (red) points using RideMood with 2$^{nd}$-order GC filter. Bottom: $\overline{mood}(t,5)$ (green when $\overline{mood}(t,5) > 0$ and red when $\overline{mood}(t,5) < 0$).

---

[1] The stock price data were downloaded from http://finance.yahoo.com and were adjusted for dividends and splits.

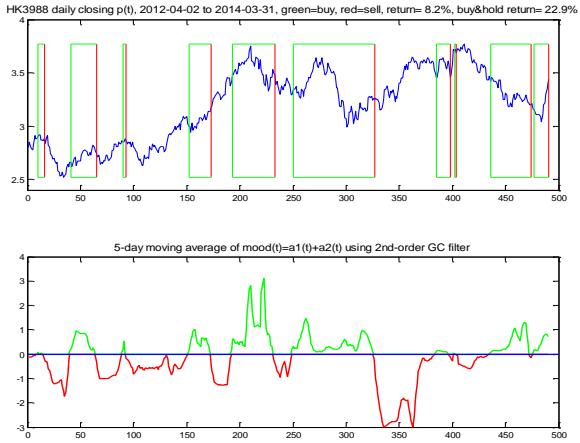

Fig. 7.5: Top: HK3988 daily closing $p_t$ from 2012-04-04 to 2014-03-31 and buy (green), sell (red) points using RideMood with 2$^{nd}$-order GC filter. Bottom: $\overline{mood}(t,5)$ (green when $\overline{mood}(t,5) > 0$ and red when $\overline{mood}(t,5) < 0$).

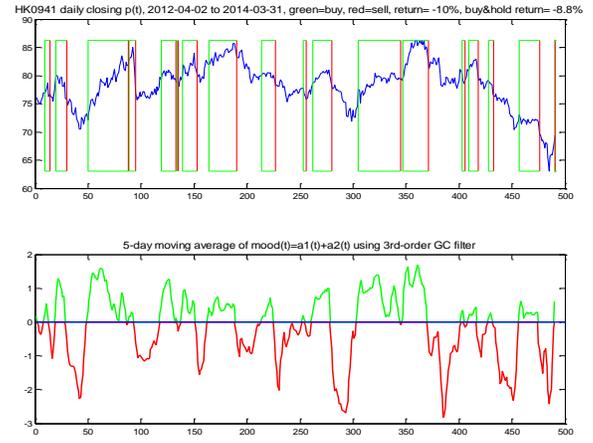

Fig. 8.3: Top: HK0941 daily closing $p_t$ from 2012-04-04 to 2014-03-31 and buy (green), sell (red) points using RideMood with 3$^{rd}$-order GC filter. Bottom: $\overline{mood}(t,5)$ (green when $\overline{mood}(t,5) > 0$ and red when $\overline{mood}(t,5) < 0$).

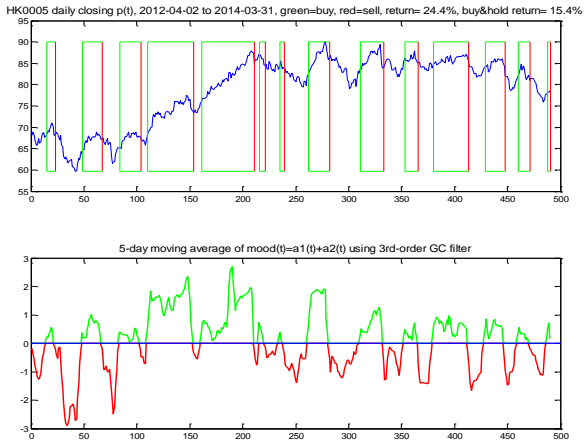

Fig. 8.1: Top: HK0005 daily closing $p_t$ from 2012-04-04 to 2014-03-31 and buy (green), sell (red) points using RideMood with 3$^{rd}$-order GC filter. Bottom: $\overline{mood}(t,5)$ (green when $\overline{mood}(t,5) > 0$ and red when $\overline{mood}(t,5) < 0$).

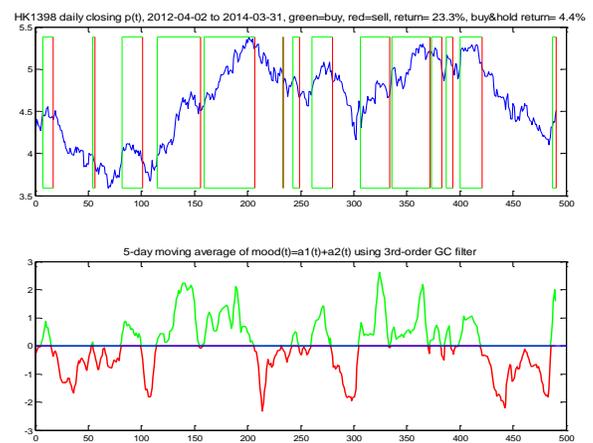

Fig. 8.4: Top: HK1398 daily closing $p_t$ from 2012-04-04 to 2014-03-31 and buy (green), sell (red) points using RideMood with 3$^{rd}$-order GC filter. Bottom: $\overline{mood}(t,5)$ (green when $\overline{mood}(t,5) > 0$ and red when $\overline{mood}(t,5) < 0$).

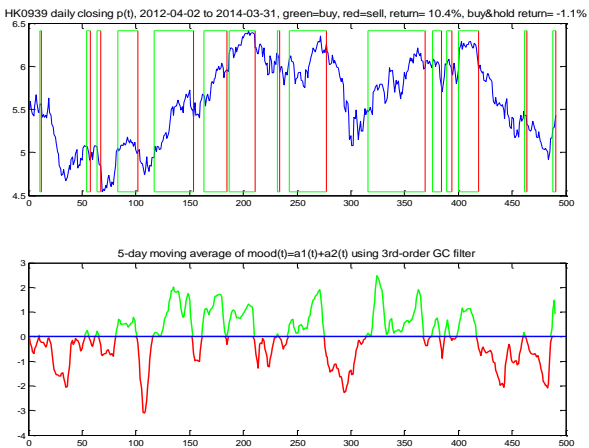

Fig. 8.2: Top: HK0939 daily closing $p_t$ from 2012-04-04 to 2014-03-31 and buy (green), sell (red) points using RideMood with 3$^{rd}$-order GC filter. Bottom: $\overline{mood}(t,5)$ (green when $\overline{mood}(t,5) > 0$ and red when $\overline{mood}(t,5) < 0$).

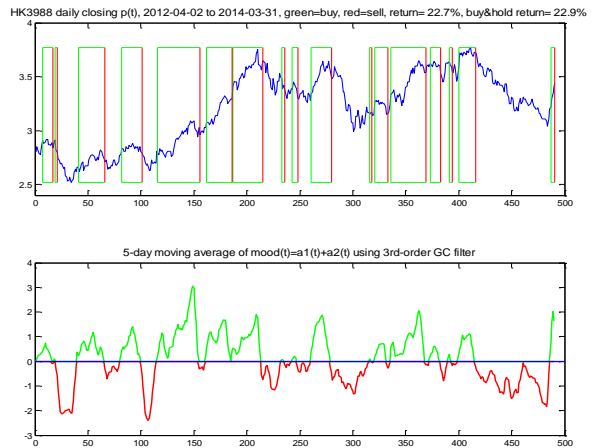

Fig. 8.5: Top: HK3988 daily closing $p_t$ from 2012-04-04 to 2014-03-31 and buy (green), sell (red) points using RideMood with 3$^{rd}$-order GC filter. Bottom: $\overline{mood}(t,5)$ (green when $\overline{mood}(t,5) > 0$ and red when $\overline{mood}(t,5) < 0$).

**Table 2: Details of the buy-sell cycles (buy/sell: price; date; return) using the 2$^{nd}$-order GC filter over the two years from 2012-04-02 to 2014-03-31 for HK0005, HK0939, HK0941, HK1398 and HK3988.**

| Cycle No. | HK0005 | HK0939 | HK0941 | HK1398 | HK3988 |
|---|---|---|---|---|---|
| 1 | Buy: 70.98; 2012-05-03<br>Sell: 68.11; 2012-05-07<br>Return: -4.04% | Buy: 4.86; 2012-06-11<br>Sell: 4.91; 2012-06-25<br>Return: 1.03% | Buy: 77.48; 2012-04-19<br>Sell: 74.27; 2012-05-29<br>Return: -4.14% | Buy: 4; 2012-09-18<br>Sell: 4.07; 2012-09-27<br>Return: 1.75% | Buy: 2.9; 2012-04-18<br>Sell: 2.89; 2012-04-26<br>Return: -0.34% |
| 2 | Buy: 60.36; 2012-06-01<br>Sell: 66.56; 2012-07-12<br>Return: 10.27% | Buy: 5.2; 2012-10-04<br>Sell: 5.95; 2012-12-27<br>Return: 14.42% | Buy: 74.87; 2012-06-18<br>Sell: 77.21; 2012-08-28<br>Return: 3.12% | Buy: 4.13; 2012-10-04<br>Sell: 5.27; 2013-01-30<br>Return: 27.6% | Buy: 2.66; 2012-06-01<br>Sell: 2.74; 2012-07-06<br>Return: 3% |
| 3 | Buy: 65.12; 2012-07-31<br>Sell: 73.54; 2012-11-15<br>Return: 12.93% | Buy: 6.15; 2013-01-02<br>Sell: 6.24; 2013-01-17<br>Return: 1.46% | Buy: 77.27; 2012-09-17<br>Sell: 79.38; 2012-10-10<br>Return: 2.73% | Buy: 4.88; 2013-05-02<br>Sell: 4.86; 2013-06-03<br>Return: -0.4% | Buy: 2.85; 2012-08-10<br>Sell: 2.86; 2012-08-15<br>Return: 0.35% |
| 4 | Buy: 76.12; 2012-11-22<br>Sell: 84.48; 2013-02-22<br>Return: 10.98% | Buy: 6.33; 2013-01-21<br>Sell: 5.93; 2013-03-05<br>Return: -6.31% | Buy: 81.77; 2012-11-05<br>Sell: 80.27; 2012-11-12<br>Return: -1.83% | Buy: 4.31; 2013-07-03<br>Sell: 5; 2013-12-16<br>Return: 16% | Buy: 3; 2012-11-09<br>Sell: 3.18; 2012-12-10<br>Return: 6% |
| 5 | Buy: 80.77; 2013-04-15<br>Sell: 80.72; 2013-04-22<br>Return: -0.06% | Buy: 5.93; 2013-04-12<br>Sell: 5.96; 2013-04-19<br>Return: 0.5% | Buy: 84.11; 2012-12-05<br>Sell: 84.21; 2013-01-10<br>Return: 0.11% | Buy: 4.37; 2014-03-27<br>Sell: 4.51; 2014-03-31<br>Return: 3.2% | Buy: 3.48; 2013-01-10<br>Sell: 3.48; 2013-03-12<br>Return: 0% |
| 6 | Buy: 82.51; 2013-04-25<br>Sell: 86.8; 2013-05-28<br>Return: 5.2% | Buy: 6.21; 2013-04-30<br>Sell: 6.17; 2013-05-21<br>Return: -0.64% | Buy: 79.9; 2013-02-07<br>Sell: 78.91; 2013-03-08<br>Return: -1.23% | | Buy: 3.34; 2013-04-09<br>Sell: 3.25; 2013-07-31<br>Return: -2.69% |
| 7 | Buy: 85.9; 2013-05-31<br>Sell: 83.45; 2013-08-22<br>Return: -2.85% | Buy: 5.24; 2013-07-04<br>Sell: 5.25; 2013-07-09<br>Return: 0.19% | Buy: 78.82; 2013-03-12<br>Sell: 75.91; 2013-03-18<br>Return: -3.69% | | Buy: 3.45; 2013-10-25<br>Sell: 3.46; 2013-11-13<br>Return: 0.29% |
| 8 | Buy: 83.75; 2013-09-03<br>Sell: 84.25; 2013-09-30<br>Return: 0.59% | Buy: 5.72; 2013-07-29<br>Sell: 5.57; 2013-08-07<br>Return: -2.62% | Buy: 76.76; 2013-03-21<br>Sell: 77.04; 2013-03-28<br>Return: 0.36% | | Buy: 3.7; 2013-11-19<br>Sell: 3.69; 2013-11-21<br>Return: -0.27% |
| 9 | Buy: 84.8; 2013-10-16<br>Sell: 84.2; 2013-12-09<br>Return: -0.7% | Buy: 5.95; 2013-08-19<br>Sell: 5.7; 2013-08-29<br>Return: -4.2% | Buy: 76.8; 2013-04-15<br>Sell: 77.88; 2013-05-27<br>Return: 1.41% | | Buy: 3.42; 2014-01-09<br>Sell: 3.2; 2014-03-06<br>Return: -6.43% |
| 10 | Buy: 81.8; 2013-12-16<br>Sell: 82; 2013-12-20<br>Return: 0.24% | Buy: 5.93; 2013-09-03<br>Sell: 5.98; 2013-10-02<br>Return: 0.84% | Buy: 77.25; 2013-07-05<br>Sell: 79.65; 2013-08-05<br>Return: 3.1% | | Buy: 3.16; 2014-03-11<br>Sell: 3.44; 2014-03-31<br>Return: 8.86% |
| 11 | Buy: 83.5; 2013-12-24<br>Sell: 82.1; 2014-01-27<br>Return: -1.67% | Buy: 6.03; 2013-10-09<br>Sell: 5.85; 2013-12-31<br>Return: -2.98% | Buy: 80.56; 2013-08-15<br>Sell: 79.12; 2013-08-28<br>Return: -1.78% | | |
| 12 | Buy: 84.85; 2014-02-19<br>Sell: 78.55; 2014-03-31<br>Return: -7.42% | | Buy: 79.84; 2013-08-30<br>Sell: 78.57; 2013-11-05<br>Return: -1.59% | | |
| 13 | | | Buy: 79.45; 2013-11-07<br>Sell: 77.79; 2013-11-13<br>Return: -2.08% | | |
| 14 | | | Buy: 80.62; 2013-11-19<br>Sell: 78.86; 2013-12-17<br>Return: -2.18% | | |
| 15 | | | Buy: 78.57; 2013-12-18<br>Sell: 75.79; 2014-01-14<br>Return: -3.53% | | |
| Accumulated Return | 23.5% | 0.3% | -11% | 54.8% | 8.2% |
| Buy&Hold Return | 15.4% | -1.1% | -8.8% | 4.4% | 22.9% |

Table 3: Details of the buy-sell cycles (buy/sell: price; date; return) using the 3rd-order GC filter over the two years from 2012-04-02 to 2014-03-31 for HK0005, HK0939, HK0941, HK1398 and HK3988.

| Cycle No. | HK0005 | HK0939 | HK0941 | HK1398 | HK3988 |
|---|---|---|---|---|---|
| 1 | Buy: 68.46; 2012-04-25<br>Sell: 68.41; 2012-05-08<br>Return: -0.07% | Buy: 5.56; 2012-04-19<br>Sell: 5.56; 2012-04-20<br>Return: 0% | Buy: 76.85; 2012-04-17<br>Sell: 76.72; 2012-04-24<br>Return: -0.16% | Buy: 4.49; 2012-04-13<br>Sell: 4.4; 2012-04-27<br>Return: -2% | Buy: 2.91; 2012-04-13<br>Sell: 2.84; 2012-04-27<br>Return: -2.4% |
| 2 | Buy: 64.87; 2012-06-13<br>Sell: 67.21; 2012-07-10<br>Return: 3.6% | Buy: 5.07; 2012-06-20<br>Sell: 4.92; 2012-06-26<br>Return: -2.95% | Buy: 80.09; 2012-05-03<br>Sell: 75.77; 2012-05-17<br>Return: -5.39% | Buy: 3.99; 2012-06-20<br>Sell: 3.88; 2012-06-22<br>Return: -2.75% | Buy: 2.91; 2012-05-02<br>Sell: 2.8; 2012-05-04<br>Return: -3.78% |
| 3 | Buy: 65.12; 2012-08-02<br>Sell: 67.19; 2012-08-30<br>Return: 3.17% | Buy: 5.06; 2012-07-05<br>Sell: 4.85; 2012-07-10<br>Return: -4.15% | Buy: 72.12; 2012-06-14<br>Sell: 81.9; 2012-08-08<br>Return: 13.56% | Buy: 4.01; 2012-07-31<br>Sell: 3.88; 2012-08-27<br>Return: -3.24% | Buy: 2.66; 2012-06-01<br>Sell: 2.71; 2012-07-09<br>Return: 1.88% |
| 4 | Buy: 68.44; 2012-09-07<br>Sell: 73.24; 2012-11-13<br>Return: 7.01% | Buy: 5.05; 2012-08-01<br>Sell: 5.01; 2012-08-28<br>Return: -0.79% | Buy: 82.82; 2012-08-09<br>Sell: 76.85; 2012-08-17<br>Return: -7.2% | Buy: 3.99; 2012-09-14<br>Sell: 4.51; 2012-11-15<br>Return: 13.03% | Buy: 2.83; 2012-07-31<br>Sell: 2.79; 2012-08-27<br>Return: -1.41% |
| 5 | Buy: 76.12; 2012-11-22<br>Sell: 84.97; 2013-02-05<br>Return: 11.62% | Buy: 4.94; 2012-09-18<br>Sell: 5.41; 2011-11-13<br>Return: 9.51% | Buy: 79.29; 2012-09-20<br>Sell: 79.05; 2012-10-12<br>Return: -0.3% | Buy: 4.56; 2012-11-20<br>Sell: 5.27; 2013-01-30<br>Return: 15.57% | Buy: 2.79; 2012-09-17<br>Sell: 2.96; 2012-11-15<br>Return: 6.09% |
| 6 | Buy: 87.06; 2013-02-15<br>Sell: 84.48; 2013-02-22<br>Return: -2.96% | Buy: 5.66; 2012-11-26<br>Sell: 5.96; 2012-12-28<br>Return: 5.3% | Buy: 78.82; 2012-10-15<br>Sell: 78.44; 2012-10-16<br>Return: -0.48% | Buy: 4.94; 2013-03-12<br>Sell: 4.84; 2013-03-13<br>Return: -2.02% | Buy: 3.08; 2012-11-23<br>Sell: 3.29; 2012-12-31<br>Return: 6.81% |
| 7 | Buy: 84.28; 2013-03-14<br>Sell: 83.76; 2013-03-20<br>Return: -0.61% | Buy: 6.15; 2013-01-02<br>Sell: 6.21; 2013-02-05<br>Return: 0.97% | Buy: 80.08; 2012-10-22<br>Sell: 80.27; 2012-11-12<br>Return: 0.23% | Buy: 4.87; 2013-03-26<br>Sell: 4.66; 2013-04-08<br>Return: -4.31% | Buy: 3.41; 2013-01-02<br>Sell: 3.65; 2013-02-24<br>Return: 7.03% |
| 8 | Buy: 82.51; 2013-04-25<br>Sell: 86.4; 2013-05-27<br>Return: 4.71% | Buy: 6.15; 2013-03-11<br>Sell: 5.93; 2013-03-13<br>Return: -3.57% | Buy: 81.96; 2012-11-27<br>Sell: 83.46; 2013-01-07<br>Return: 1.83% | Buy: 4.76; 2013-04-24<br>Sell: 4.85; 2013-05-23<br>Return: 1.89% | Buy: 3.48; 2013-03-12<br>Sell: 3.38; 2013-03-15<br>Return: -2.87% |
| 9 | Buy: 83.7; 2013-07-09<br>Sell: 84..5; 2013-08-08<br>Return: 0.41% | Buy: 6.08; 2013-03-26<br>Sell: 6.26; 2013-05-20<br>Return: 2.96% | Buy: 0.08; 2013-02-08<br>Sell: 78.26; 2013-03-04<br>Return: -2.27% | Buy: 4.49; 2013-07-02<br>Sell: 4.7; 2013-08-09<br>Return: 4.67% | Buy: 3.38; 2013-03-26<br>Sell: 3.3; 2013-04-05<br>Return: -2.36% |
| 10 | Buy: 86.25; 2013-09-06<br>Sell: 85.1; 2013-09-26<br>Return: -1.33% | Buy: 5.47; 2013-07-16<br>Sell: 5.98; 2013-10-02<br>Return: 9.32% | Buy: 77.74; 2013-04-12<br>Sell: 75.82; 2013-04-17<br>Return: -2.46% | Buy: 4.99; 2013-08-13<br>Sell: 5.17; 2013-10-07<br>Return: 3.6% | Buy: 3.34; 2013-04-24<br>Sell: 3.48; 2013-05-23<br>Return: 4.19% |
| 11 | Buy: 84.75; 2013-10-18<br>Sell: 85.4; 2013-12-04<br>Return: 0.76% | Buy: 6.04; 2013-10-11<br>Sell: 5.76; 2013-10-24<br>Return: -4.63% | Buy: 78.68; 2013-04-25<br>Sell: 78.12; 2013-05-23<br>Return: -0.71% | Buy: 5.21; 2013-10-08<br>Sell: 4.99; 2013-10-23<br>Return: -4.22% | Buy: 3.15; 2013-07-16<br>Sell: 3.17; 2013-07-18<br>Return: 0.64% |
| 12 | Buy: 84.15; 2013-12-30<br>Sell: 82.1; 2014-01-27<br>Return: -2.43% | Buy: 6.02; 2013-10-31<br>Sell: 5.93; 2013-11-07<br>Return: -1.49% | Buy: 77.59; 2013-06-28<br>Sell: 79.46; 2013-08-27<br>Return: 2.41% | Buy: 5.05; 2013-10-29<br>Sell: 5.03; 2013-11-06<br>Return: -0.39% | Buy: 3.27; 2013-07-23<br>Sell: 3.16; 2013-08-08<br>Return: -3.36% |
| 13 | Buy: 82.1; 2014-02-14<br>Sell: 81.05; 2014-03-03<br>Return: -1.27% | Buy: 5.97; 2013-11-15<br>Sell: 5.97; 2013-12-12<br>Return: 0% | Buy: 79.6; 2013-08-29<br>Sell: 82.09; 2013-10-04<br>Return: 3.12% | Buy: 5.04; 2013-11-15<br>Sell: 5; 2013-12-16<br>Return: -0.79% | Buy: 3.34; 2013-08-13<br>Sell: 3.54; 2013-10-02<br>Return: 5.98% |
| 14 | Buy: 78.25; 2014-03-26<br>Sell: 78.55; 2014-03-31<br>Return: 0.38% | Buy: 5.43; 2014-02-18<br>Sell: 5.32; 2014-02-20<br>Return: -2.02% | Buy: 80.18; 2013-11-20<br>Sell: 79.21; 2013-11-25<br>Return: -1.2% | Buy: 4.32; 2014-03-25<br>Sell: 4.51; 2014-03-31<br>Return: 4.39% | Buy: 3.62; 2013-10-08<br>Sell: 3.53; 2013-10-23<br>Return: -2.48% |
| 15 | | Buy: 5.28; 2014-03-26<br>Sell: 5.43; 2014-03-31<br>Return: 2.84% | Buy: 80.72; 2013-11-28<br>Sell: 79.55; 2013-12-11<br>Return: -1.44% | | Buy: 3.64; 2013-11-04<br>Sell: 3.55; 2013-11-07<br>Return: -2.47% |
| 16 | | | Buy: 79.16; 2013-12-27<br>Sell: 76.37; 2014-01-06<br>Return: -3.52% | | Buy: 3.59; 2013-11-15<br>Sell: 3.72; 2013-12-09<br>Return: 3.62% |
| 17 | | | Buy: 73; 2014-02-11<br>Sell: 69.43; 2014-03-10<br>Return: -4.89% | | Buy: 3.22; 2014-03-25<br>Sell: 3.44; 2014-03-31<br>Return: 6.82% |
| 18 | | | Buy: 69.38; 2014-03-31<br>Sell: 69.38; 2014-01-31<br>Return: 0% | | |
| Accumulated Return | 24.4% | 10.4% | -10% | 23.3% | 22.7% |
| Buy&Hold Return | 15.4% | -1.1% | -8.8% | 4.4% | 22.9% |

## VI. CONCLUDING REMARKS

The Gaussian-Chain distribution proposed in this paper provides a new mathematical framework to study the hierarchical structures that are prevalent in social organizations. A key insight gained from the mathematical analysis (Lemma 2) is that the mean and variance, that are usually the only variables used in standard tools, do not change with the number of levels in the hierarchy, but the kurtosis increases exponentially with the number of hierarchical levels $q$ according to $3^q - 3$. This sends the following message: the increase of hierarchy makes the "average" more average and the "extremists" more extreme (i.e. destroying the "middle class"), and, more importantly, if we look at only the low-order averages --- mean and variance, the two opposite moves cancel each other and we do not see any changes. This shows the importance of incorporating higher order statistics (and heavy-tailed distributions) into the models when we study social systems (such as stock markets).

The $2^{nd}$ and $3^{rd}$-order GC filters proposed in this paper take the heavy-tailed distribution explicitly into consideration and show much better performance than the conventional least-squares algorithm when the noise is heavy-tail distributed. The RideMood strategy with these GC filters performed much better than the benchmark Buy-and-Hold strategy and the Index Fund: for the test on five blue-chip Hong Kong stocks over the two-year period from April 2, 2012 to March 31, 2014, the average returns of the RideMood with $2^{nd}$-order GC filter, RideMood with $3^{rd}$-order GC filter, the Buy-and-Hold, and the HSI are 15.16%, 14.16%, 5.56%, and 7.76%, respectively.

## APPENDIX

Deviation of the $3^{rd}$-order Gaussian-Chain Filter: Substituting $\sigma\varepsilon_j^{(3)} = r_j - a^T ed(x_{j-1})$ into the joint density function of $\sigma\varepsilon_j^{(3)} = \sigma\varepsilon_{0,1}^{(3)} = \varepsilon_{0,\sigma}^{(3)}$ $(j = 1,2,...,t)$, we get

$$f(r_1, x_0, ..., r_t, x_{t-1}) =$$
$$\int_{-\infty}^{\infty}\cdots\int_{-\infty}^{\infty}\prod_{j=1}^{t}\frac{1}{|v_j|\sqrt{2\pi}}e^{-\frac{1}{2}\left(\frac{r_j-a^T ed(x_{j-1})}{v_j}\right)^2}\frac{1}{|u_j|\sqrt{2\pi}}e^{-\frac{1}{2}\left(\frac{v_j}{u_j}\right)^2}\frac{1}{\sigma\sqrt{2\pi}}e^{-\frac{1}{2}\left(\frac{u_j}{\sigma}\right)^2}du_j dv_j \quad (A1)$$

which is the integration of the $2t$-dimensional function

$$g(u_1,...,u_t,v_1,...,v_t) =$$
$$\prod_{j=1}^{t}\left(\frac{1}{|v_j|\sqrt{2\pi}}e^{-\frac{1}{2}\left(\frac{r_j-a^T ed(x_{j-1})}{v_j}\right)^2}\frac{1}{|u_j|\sqrt{2\pi}}e^{-\frac{1}{2}\left(\frac{v_j}{u_j}\right)^2}\frac{1}{\sigma\sqrt{2\pi}}e^{-\frac{1}{2}\left(\frac{u_j}{\sigma}\right)^2}\right) \quad (A2)$$

over $R^{2t}$. Based on the same argument as in the derivation of the $2^{nd}$-order GC filter, we define the likelihood function

$$L(a,\sigma,u_1,...,u_t,v_1,...,v_t) =$$
$$\prod_{j=1}^{t}\left(\frac{1}{|v_j|\sqrt{2\pi}}e^{-\frac{1}{2}\left(\frac{r_j-a^T ed(x_{j-1})}{v_j}\right)^2}\frac{1}{|u_j|\sqrt{2\pi}}e^{-\frac{1}{2}\left(\frac{v_j}{u_j}\right)^2}\frac{1}{\sigma\sqrt{2\pi}}e^{-\frac{1}{2}\left(\frac{u_j}{\sigma}\right)^2}\right)^{\lambda^{t-j}}$$

$$= \frac{e^{-\frac{1}{2}\sum_{j=1}^{t}\left[\left(\frac{r_j-a^T ed(x_{j-1})}{v_j}\right)^2+\left(\frac{v_j}{u_j}\right)^2+\left(\frac{u_j}{\sigma}\right)^2\right]\lambda^{t-j}}}{\prod_{j=1}^{t}(|u_j||v_j|(2\pi)^{3/2}\sigma)^{\lambda^{t-j}}} \quad (A3)$$

where $\lambda \in (0,1)$ is a weighting factor. Taking the partial derivative of the likelihood function (A3) with respect to the parameter vector $a$ and setting it to zero:

$$\frac{\partial L}{\partial a} = L(a,\sigma,u_1,...,u_t,v_1,...,v_t)$$
$$\sum_{j=1}^{t}\left(\frac{r_j - a^T ed(x_{j-1})}{v_j^2}\right)\lambda^{t-j}ed(x_{j-1}) = 0 \quad (A4)$$

we get

$$\sum_{j=1}^{t}\left(\frac{r_j - \hat{a}_t^T ed(x_{j-1})}{v_j^2}\right)\lambda^{t-j}ed(x_{j-1}) = 0 \quad (A5)$$

which is the same as (31) in the derivation of the $2^{nd}$-order Gaussian-Chain filter. Therefore following the same from (31) to (37), we obtain (47) and (48).

To find the maximum likelihood estimate of $\sigma$, take the partial derivative of the likelihood function (A3) with respect to $\sigma$ and set it to zero:

$$\frac{\partial L}{\partial \sigma} = L(a,\sigma,u_1,...,u_t,v_1,...,v_t)$$
$$\left(\frac{\sum_{j=1}^{t}u_j^2\lambda^{t-j}}{\sigma^3} - \frac{\sum_{j=1}^{t}\lambda^{t-j}}{\sigma}\right) = 0 \quad (A6)$$

which is the same as (38) in the derivation of the $2^{nd}$-order GC filter except the $v_j^2$ in (38) becomes $u_j^2$ in (A6). Therefore following the same from (38) to (41) we get (52) and (53).

Finally, we find the maximum likelihood estimate of $v_t$ and $u_t$. Rewriting the likelihood function (A3) as

$$L = [L(a,\sigma,u_1,...,u_{t-1},v_1,...,v_{t-1})]^{\lambda}$$
$$\left(\frac{e^{-\frac{1}{2}\left[\left(\frac{r_t-a^T ed(x_{t-1})}{v_t}\right)^2+\left(\frac{v_t}{u_t}\right)^2+\left(\frac{u_t}{\sigma}\right)^2\right]}}{|u_t||v_t|(2\pi)^{3/2}\sigma}\right) \quad (A7)$$

we see that finding $v_t$ and $u_t$ to maximize $L(a,\sigma,u_1,...,u_t,v_1,...,v_t)$ is equivalent to maximizing $\bar{P}(r_t,x_{t-1}|a,\sigma,u_t,v_t) \equiv$

$$\frac{1}{|u_t||v_t|(2\pi)^{3/2}\sigma}e^{-\frac{1}{2}\left[\left(\frac{r_t-a^T ed(x_{t-1})}{v_t}\right)^2+\left(\frac{v_t}{u_t}\right)^2+\left(\frac{u_t}{\sigma}\right)^2\right]} \quad (A8)$$

Taking the partial derivatives of $\bar{P}(r_t,x_t|a,\sigma,u_t,v_t)$ with respect to $|v_t|$ and $|u_t|$ and setting them to zero:

$$\frac{\partial \bar{P}(r_t,x_{t-1}|a,\sigma,u_t,v_t)}{\partial |v_t|} =$$
$$\bar{P}(r_t,x_{t-1}|a,\sigma,u_t,v_t)\left[\frac{(r_t - a^T ed(x_{t-1}))^2}{|v_t|^3} - \frac{|v_t|}{u_t^2} - \frac{1}{|v_t|}\right] = 0 \quad (A9)$$

$$\frac{\partial \bar{P}(r_t,x_{t-1}|a,\sigma,u_t,v_t)}{\partial |u_t|} =$$
$$\bar{P}(r_t,x_{t-1}|a,\sigma,u_t,v_t)\left(\frac{v_t^2}{|u_t|^3} - \frac{|u_t|}{\sigma^2} - \frac{1}{|u_t|}\right) = 0 \quad (A10)$$

we get

$$|v_t|^4 + u_t^2|v_t|^2 - u_t^2\left(r_t - \hat{a}_{t-1}^T ed(x_{t-1})\right)^2 = 0 \quad (A11)$$

$$|u_t|^4 + \hat{\sigma}_{t-1}^2|u_t|^2 - \hat{\sigma}_{t-1}^2 v_t^2 = 0 \quad (A12)$$

where we replace the $a$ and $\sigma$ by their best available estimates $\hat{a}_{t-1}$ and $\hat{\sigma}_{t-1}$, respectively. (A12) gives

$$v_t^2 = \frac{u_t^4 + \hat{\sigma}_{t-1}^2 u_t^2}{\hat{\sigma}_{t-1}^2} \quad (A13)$$

which is (49), and substitute it into (A11) to get

$$(u_t^2)^3 + 3\hat{\sigma}_{t-1}^2(u_t^2)^2 + 2\hat{\sigma}_{t-1}^4 u_t^2$$
$$-\hat{\sigma}_{t-1}^4\left(r_t - \hat{a}_{t-1}^T ed(x_{t-1})\right)^2 = 0 \quad (A14)$$

whose solution is

$$u_t^2 = \sqrt[3]{\frac{s}{2} + \sqrt{\left(\frac{s}{2}\right)^2 - \frac{\hat{\sigma}_{t-1}^{12}}{27}}} + \sqrt[3]{\frac{s}{2} - \sqrt{\left(\frac{s}{2}\right)^2 - \frac{\hat{\sigma}_{t-1}^{12}}{27}}} - \hat{\sigma}_{t-1}^2 \quad (A15)$$

with

$$s = \hat{\sigma}_{t-1}^4\left[\left(r_t - \hat{a}_{t-1}^T ed(x_{t-1})\right)^2 + 4\hat{\sigma}_{t-1}^2\right] \quad (A16)$$

which are (50) and (51). ∎

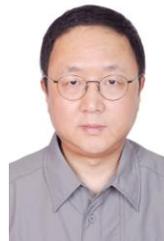

**Li-Xin Wang** received the Ph.D. degree in 1992 from the Department of Electrical Engineering – Systems, University of Southern California (won USC's Phi Kappa Phi's highest Student Recognition Award). From 1992 to 1993 he was a Postdoc Fellow with the Department of Electrical Engineering and Computer Science, University of California at Berkeley. From 1993 to 2007 he was on the faculty of the Department of Electronic and Computer Engineering, The Hong Kong University of Science and Technology. In 2007 he resigned from his tenured position at HKUST to become an independent researcher and investor in the stock and real estate markets in Hong Kong and China. He returned to academic in Fall 2013 by joining the faculty of the Department of Automation Science and Technology, Xian Jiaotong University, after a fruitful hunting journey across the wild land of investment to achieve financial freedom.

His research interests are dynamical models of asset prices, market microstructure, trading strategies, fuzzy systems, and adaptive nonlinear control.